\documentclass{aa}  
\usepackage{mydef}
\usepackage{graphicx}
\usepackage{color}
\usepackage{enumitem}
\usepackage{txfonts}
\begin{document}

   \title{Far-infrared study of tracers of oxygen chemistry in diffuse clouds}

   \author{H. Wiesemeyer \inst{1}
          \and
           R. G{\"u}sten \inst{1}
          \and
           S. Heyminck \inst{1}
          \and
           H.W. H{\"u}bers \inst{2,3}
          \and
           K.M. Menten \inst{1} 
          \and
           D.A. Neufeld\inst{4}
          \and
          H. Richter\inst{2}
          \and
          R. Simon\inst{5}
          \and
           J. Stutzki\inst{5}
          \and
           B. Winkel\inst{1}
          \and
           F. Wyrowski\inst{1}
          }

   \institute{Max-Planck-Institut f{\"u}r Radioastronomie,
              Auf dem H{\"u}gel 69, 53121 Bonn, Germany \\
              \email{hwiese@mpifr.de}
         \and
              Deutsches Zentrum f{\"u}r Luft- und Raumfahrt, 
              Institute of Optical Sensor Systems,
              Rutherfordstr. 2, 12489 Berlin, Germany
         \and
              Humboldt-Universit{\"a}t zu Berlin, Department of Physics,
              Newtonstr. 15, 12489 Berlin, Germany 
         \and
              The Johns Hopkins University, 3400 North Charles St. Baltimore,
              MD 21218, USA
         \and
              I. Physikalisches Institut, Universit{\"a}t zu K{\"o}ln,
              Z{\"u}lpicher Str. 77, 50937 K{\"o}ln, Germany
             }

   \date{Received ; accepted }

 
  \abstract
   {The chemistry of the diffuse interstellar medium rests upon three
    pillars: exothermic ion-neutral reactions (``cold chemistry''),
    endothermic neutral-neutral reactions with significant activation
    barriers (``warm chemistry''), and reactions on the surfaces of
    dust grains. While warm chemistry becomes important in the shocks
    associated with turbulent dissipation regions, the main path
    for the formation of interstellar OH and $\HHO$ is that of cold
    chemistry.}
   {The aim of this study is to observationally confirm the association of 
    atomic oxygen with both atomic and molecular gas phases, and to understand
    the measured abundances of $\OH$ and $\OHP$ as a function of the available
    reservoir of $\HH$.}
   {We obtained absorption spectra of the ground states of OH, $\OHP$ and \OI  with high-velocity resolution, with GREAT
    onboard SOFIA, and with the THz receiver at the APEX. We analyzed them along with ancillary spectra of HF and CH from HIFI.
    To deconvolve them from the hyperfine structure and to separate the blend that is due to various velocity components
    on the sightline, we fit model spectra consisting of an appropriate number of Gaussian profiles using a
    method combining simulated annealing with downhill simplex minimization. Together with HF and/or CH as a surrogate
    for $\HH$, and \HI $\lambda$21~cm data, the molecular hydrogen fraction $f^{\rm N}_\HH = N(\HH)/(N(\H)+2N(\HH))$ can be
    determined. We then investigated abundance ratios as a function of $f^{\rm N}_\HH$.
   }
   {The column density of \OI is correlated at a high significance with the amount of available molecular and atomic
    hydrogen, with an atomic oxygen abundance of $3\times 10^{-4}$ relative to H nuclei.
    While the velocities of the absorption features of OH and $\OHP$ are loosely correlated and reflect the spiral arm
    crossings on the sightline, upon closer inspection they display an anticorrespondence. The arm-to-interarm density
    contrast is found to be higher in OH than in $\OHP$. While both species can coexist, with a higher abundance in OH than
    in $\OHP$, the latter is found less frequently in absence of OH than the other way around, which is a direct consequence
    of the rapid destruction of $\OHP$ by dissociative recombination when not enough
    $\HH$ is available. This conjecture has been substantiated by a comparison between the OH/$\OHP$ ratio with
    $f^{\rm N}_\HH$, showing a clear correlation. The hydrogen abstraction reaction chain $\OHP(\HH,\H)\HHOP(\HH,\H)\HHHOP$
    is confirmed as the pathway for the production of OH and $\HHO$. Our estimate of the branching ratio of the
    dissociative recombination of $\HHHOP$ to OH and $\HHO$ is confined within the interval of 84 to 91\%, which matches
    laboratory measurements (74 to 83\%). -- A correlation between the linewidths and column densities of $\OHP$
    features is found to be significant with a false-alarm probability below 5\%. Such a correlation is predicted
    by models of interstellar MHD turbulence. For OH the same correlation is found to be insignificant because there are more narrow absorption features.}
   {While it is difficult to assess the contributions of warm neutral-neutral chemistry to the observed abundances, it seems
    fair to conclude that the predictions of cold ion-neutral chemistry match the abundance patterns we observed.}

   \maketitle
%

\section{Introduction}
The gas in the diffuse atomic and molecular phases of the interstellar medium
(ISM) contributes roughly 5\% to the total visible mass of our Galaxy
\citep{1997ApJ...480..173S,2014ApJ...794...59K} and represents the reservoir
out of which the denser gas phases are formed when crossing the spiral arms
and the Galactic bulge. While most of the molecular gas is concentrated in
molecular clouds and their star-forming cores, the discovery of molecules in the
diffuse phase \citep{1937PASP...49...26D,1937ApJ....86..483S} came as a surprise,
given the strong dissociating UV radiation in the Galaxy. An account of the
historical development is given in \citet{2005ARA&A..43..337C}. Because diffuse
clouds can be observed in spectroscopic tracers from UV to radio wavelengths,
the diffuse gas was soon shown to be chemically extremely rich
(\citealp{2006ARA&A..44..367S}, further references therein). The chemistry of
diffuse clouds has been reviewed by \citet{1998ISAA....4...53V}. The accepted
theory rests on three pillars: cold chemistry driven by ion-neutral
reactions, warm chemistry whose endothermic reactions are powered by the
dissipation of supersonic turbulence, and chemistry on surfaces of dust grains
acting as catalyst.

While the first comprehensive models of chemistry in diffuse gas considered
a realistic cloud structure and radiative transfer \citep{1992MNRAS.258..377H,1986ApJS...62..109V}
and thereby corrected the previously underestimated self-shielding against the
interstellar UV radiation field, they considered mainly cold chemistry and
accounted for grain surface chemistry only in the formation of molecular
hydrogen. Warm chemistry has been studied by \citet{2009A&A...495..847G},
making use of our best knowledge of interstellar turbulence and its
dissipation. \citet{2009ApJ...691.1459V} have demonstrated the Monte Carlo treatment of
gas-grain chemistry in large reaction networks.

The advent of low-noise Terahertz receivers with high-resolution spectrometers
such as HIFI \citep{2010A&A...518L...6D} and GREAT\footnote{GREAT is a development by the MPI f{\"u}r
Radioastronomie and the KOSMA/ Universit{\"a}t zu K{\"o}ln, in cooperation with
the MPI f{\"u}r Sonnensystemforschung and the DLR Institut f{\"u}r
Planetenforschung.} \citep{2012A&A...542L...1H} onboard the Herschel and SOFIA
observatories, respectively, offered a new opportunity to derive column
densities of light hydrides from measurements of absorption out of the ground state. Their precision
is only limited by the radiometric noise and the accuracy of the calibration of
the illuminating continuum radiation, but not fraught anymore with uncertain
assumptions regarding the excitation of the radio lines used previously.

All atomic and molecular constituents of the network
of chemical reactions leading to the formation of interstellar water and
hydroxyl are accessible to observations. The latest contribution to these studies is
the high-resolution spectroscopy of the ground-state transition of the fine
structure of atomic oxygen (\OI) with GREAT. Because of its usefulness as a tracer for the total gas mass in diffuse clouds, it allows us to understand
the abundance of interstellar water and hydroxyl in more detail than previously
possible. The latter two species are both mainly formed from the $\OHP$ cation, whose abundance
depends on the reservoir of available hydrogen and its molecular fraction,
on the cosmic ray ionization rate, and on the initially available \OI. The $\OHP$ cation
was first detected by \citet{2010A&A...518A..26W}. Motivated by the increasing amount of high-quality data,
models, and databases for chemical networks (e.g., OSU, http://www.physics.ohio-state.edu/~eric/research.html,
and KIDA, \citealp{2012EAS....58..287W} and http://kida.obs.u-bordeaux1.fr/), we investigate the
abundance patterns in the diffuse gas of the spiral arms on the sightlines to continuum sources in the first
and fourth quadrants of the Galaxy.

The plan of this work is as follows: Section \ref{sec:obs} presents the technical aspects of the
observations and the data analysis. Section \ref{sec:data} provides a digest of the data used
for this study. A phenomenological description of the results can be found in Appendix
\ref{app:A}. The discussion in Sect. \ref{sec:discussion} presents a deeper analysis of the
data, referring to Appendix \ref{app:B} for a note on averaging abundances. We conclude the paper with
Sect. \ref{sec:conclusions}. The most important chemical reactions addressed
in this work are summarized in Appendix \ref{app:C}; a few simple chemical models are presented in
Appendix \ref{app:D}. In Appendix \ref{app:E} we assess the reliability of the CH and HF ground-state transitions as tracers for $\HH$. A comparison of hydroxyl and oxygen spectroscopy with PACS
\citep{2010A&A...518L...2P} and GREAT is shown in Appendix \ref{app:F} to demonstrate
the cross-calibration of the two instruments.

\section{Observations and data analysis}
\label{sec:obs}
The following sections provide an overview of the technical aspects of our
observations with GREAT and with the terahertz receiver at the APEX
\footnote{This publication is based on data acquired with the Atacama Pathfinder
Experiment (APEX). APEX is a collaboration between the Max-Planck-Institut f{\"u}r Radioastronomie,
the European Southern Observatory, and the Onsala Space Observatory.} \citep{2010stt..conf..130L}.
Our source selection consists of sources from the Herschel guaranteed time program PRISMAS
(P.I. M.~Gerin,\footnote{http://astro.ens.fr/?PRISMAS}) and far-IR-bright hot cores in the fourth quadrant of the Galaxy
\citep[from ATLASGAL,][]{2014A&A...565A..75C}.
The set of data that we obtained was completed with HIFI spectra from the Herschel Science Archive; the underlying
observations are described elsewhere (e.g., Gerin et al. 2010, further
references therein). Table~\ref{table:1} summarizes the observed transitions
and the origin of the data.

%
\begin{table*}[ht!]
\caption{List of studied species and transitions.}
\label{table:1}      
\centering          
\begin{tabular}{l l l r r c r l}
\hline\hline       
Species       & Transition & hyperfine & Frequency & \multicolumn{1}{c}{$\Delta\upsilon_{\rm HFS}$} & $A_{\rm E}$  &
$E_{\rm U}$~~~ & Origin \\ 
              &            & component & [GHz]~~~  & [\kms] & [s$^{-1}$]  & [K]~~~
    &        \\
\hline                    
\OI    & $^3{\rm P}_1 \leftarrow \,\,^3{\rm P}_2$ & &
         4744.7775 & & $8.91\times 10^{-5}$ & 227.76 & GREAT \\
OH     & $^2\Pi_{3/2}\,\, J=5/2\leftarrow 3/2$ & $F=2^- \leftarrow 2^+$ & 
         2514.2987  & $+2.1$ & 0.0137 &  120.75      & GREAT \\
       &                                   & $F=3^- \leftarrow 2^+$ &
         2514.3167  & $0.0$  & 0.1368 &             \\ 
       &                                   & $F=2^- \leftarrow 1^+$ &
         2514.3532 & $-4.4$ & 0.1231 &       \\
HF     & $J=1\leftarrow 0$    &             & 1232.4763 & & 0.0242  & 59.15 & HIFI \\
$\OHP$ & $N= 1 \leftarrow 0,\,\, J=1\leftarrow 1$ & $F=1/2\leftarrow 1/2$ & 1032.9979 & $+35.0$& 0.0141  & 49.58 & APEX \\
       &                 & $F=3/2\leftarrow 1/2$ & 1033.0044  & $+33.1$& 0.0035  &       & \\
       &                 & $F=1/2\leftarrow 3/2$ & 1033.1118  & $+2.0$ & 0.0070  &       & \\
       &                 & $F=3/2\leftarrow 3/2$ & 1033.1186  & $0.0$  & 0.0176  &       & \\
       & $N= 1\leftarrow 0,\,\, J=2\leftarrow 1$ & $F=5/2\leftarrow 3/2$ &  971.8038  & $0.0$ & 0.0182 & 46.64 & HIFI \\
       &                 & $F=3/2\leftarrow 1/2$ &  971.8053  & $-0.46$ & 0.0152  &       & \\
       &                 & $F=3/2\leftarrow 3/2$ &  971.9192  & $-35.6$ & 0.0030  &       & \\
CH     & $N=1,\,\, J=3/2\leftarrow 1/2$& $F=2^-\leftarrow 1^+$ &  536.7611 & $19.3$ & 0.0006 & 25.76 & HIFI \\
       &                 & $F=1^-\leftarrow 1^+$ &  536.7819  & $7.7$ & 0.0002  &       &      \\
       &                 & $F=1^-\leftarrow 0^+$ &  536.7957  & $0.0$ & 0.0004  &       &      \\
\hline                  
\end{tabular}
\tablefoot{Frequencies, Einstein coefficients, and upper level energies are
from the JPL \citep{1998JQSRT..60..883P} and CDMS \citep{2001A&A...370L..49M} databases.}
\end{table*}
%
\begin{table*}
\caption{Continuum sources and observed species.}             
\label{table:3}      
\centering          
\begin{tabular}{l c c r r r c c l }
\hline\hline       
    & $T_{\rm c}^{(a)}$ [K] & $\alpha$ (J2000) & $\delta$ (J2000) & \multicolumn{1}{c}{l}  & \multicolumn{1}{c}{b} & \multicolumn{2}{c}{\vlsr [\kms]$^{(b)}$} & Species \\ 
\hline                    
G10.47        & 6.9 & 18:08:38.20 & $-$19:51:50.0 & 10\fdg472 & 0\fdg027     & $(+48,+69)$ & $(+58,+77)$  &
 CH, OH, $\OHP$  \\  
G10.62$^{(c)}$  & 10.4 & 18:10:28.69 & $-$19:55:50.0 & 10.621 & $-0.387$  & $(-3,+1)$ & $(-10,+7)$   & 
 CH, HF, \OI, OH, $\OHP$ \\
G34.26        & 9.0 & 18:53:18.70 &  01:14:58.0 &  34.257 & 0.154      & $(+55,+62)$ & $(+58,+62)$  &
 HF, \OI, OH, $\OHP$  \\
W49N          & 12.3 & 19:10:13.20 &  09:06:12.0 &  43.166 & 0.012      & $(+2,+21)$ & $(-1,+19)$   &
 CH, HF, \OI, OH, $\OHP$  \\
W51e2         & 8.0 & 19:23:43.90 &  14:30:30.5 &  49.489 & $-0.388$   & $(+47,+72)$ & $(+50,+61)$  &
 CH, HF, OH, $\OHP$ \\
G327.29      & 3.5 & 15:53:08.55 & $-$54:37:05.1 & 327.294 & $-0.580$  & $(-72,-40)$ & $(-49,-37)$  &
 CH, OH, $\OHP$  \\
G330.95       & 10.4 & 16:09:53.01 & $-$51:54:55.0 & 330.954 & $-0.182$ & $(-102,-80)$ & $(-91,-87)$  &
 CH, OH, $\OHP$ \\
G332.83       & 7.1 & 16:20:10.65 & $-$50:53:17.6 & 332.824 & $-0.548$ & $(-55,-53)$ & $(-59,-45)$  &
 OH, $\OHP$   \\
G351.58       & 3.2 & 17:25:25.03 & $-$36:12:45.3 & 351.581 & $-0.353$ & $(-101,-89)$ & $(-100,-87)$ &
 OH, $\OHP$   \\
\hline                  
\end{tabular}
\tablefoot{$^{(a)}$ Main-beam brightness temperature of continuum at 2.514~THz (Rayleigh-Jeans scale).
$^{(b)}$ Columns 6 and 7 list the velocity ranges with OH and CH$_3$OH maser emission, respectively.
For the fourth quadrant, OH and CH$_3$OH maser velocities are from \citet{1998MNRAS.297..215C} and
\citet{1995MNRAS.272...96C}, for G34.26, from \citet{2005ApJS..160..220F} and \citet{1995MNRAS.272...96C},
for W49N, from \citet{2013ApJ...775...36D} and \citet{2014A&A...564A.110B}, and for W51e2 from
\citet{2005ApJS..160..220F} and \citet{2012A&A...541A..47S}, respectively. $^{(c)}$ Also known as W31C.}
\end{table*}

\subsection{Terahertz spectroscopy of OH and \OI with GREAT}
We conducted the spectroscopy of the $^2\Pi_{3/2}\,J=5/2 \leftarrow 3/2$ ground-state line and of the excited
$^2\Pi_{1/2}\,J=5/2 \leftarrow 3/2$ line of OH with the M$_a$ channel of GREAT onboard SOFIA. To remove instrumental and
atmospheric fluctuations and to suppress standing waves in the spectra, the double beam-switch mode was used, chopping with
the secondary mirror at a rate of 1~Hz and with an amplitude of typically $60''$. The new data, mainly from the fourth quadrant,
were obtained on the southern deployment flights from New Zealand in July 2013, as a part of the observatory's cycle 1.
Pointing and tracking were accurate to better than $2''$.
The sightlines toward G10.47 and W31C were observed in April 2013, on northern hemisphere flights of the same cycle.
W49N, W51e2 and G34.26 were already observed in 2012 on the basic science flights \citep{2012A&A...542L...7W}.
With the successful commissioning of GREAT's H channel in May 2014, the fine-structure line of atomic oxygen became
accessible for spectrally resolved studies. A novel NbN HEB waveguide mixer \citep{2015ITTST...5..207B} was pumped by
a quantum cascade laser as local oscillator \citep{2015IEEE...subm}; the Doppler correction was applied offline.

The southern deployment flights offered excellent observing conditions. Typical system temperatures range between
6340 and 8300~K (single sideband scale), with a median system temperature of 6500~K, for water vapor columns between 4.3 and 14.2~$\mu$m,
with a median of 10.9~$\mu$m water vapor. The weather conditions on the northern flights in April 2013
were slightly less favorable but still very good, with system temperatures and water vapor columns of
6640~K and 15~$\mu$m, respectively (median values). In May 2014 the oxygen line was observed at a system temperature of 2400~K,
with a water vapor column of 6.6~$\mu$m (median values). The calibration of the H-channel and in the low-frequency L$_2$ 
channel is consistent and yielded the same water vapor column.

In 2013 the main-beam efficiencies were significantly improved thanks to new optics and a smaller scatter cone.
They were measured by means of total power cross-scans on Jupiter in April 2013, yielding a value of $\eta_{\rm mb} = 0.70$
for the M$_a$ channel. The major and minor sizes of the Jupiter disk were $34\farcs5$ and $32\farcs2$, respectively. For the
southern hemisphere campaign we observed Saturn as calibrator, confirming this efficiency. The beam width is $12''$ (FWHM). For
the H-channel, the main beam efficiency is 0.66, with a $6\farcs 6$ beam width. These values were determined on Mars, which had
a diameter of $13''$.

For the spectral analysis of our data we used the XFFTS spectrometer \citep{2012A&A...542L...3K}. The spectral resolution of
76.3~kHz was smoothed to a velocity resolution of 0.36~\kms (in a few cases 0.72~\kms).

\subsection{$\OHP$ spectroscopy}
The ground-state line of $\OHP$ was observed in July 2010 at the APEX telescope
with the MPIfR dual-channel terahertz receiver \citep{2010stt..conf..130L}, with the exception of
G34.26, W49N and W51e2, whose $\OHP$ HIFI spectra were retrieved from the
Herschel Science Archive. Details of the instrumental set-up at the APEX are given in \citet{2011ApJ...743L..25Q}.

The weather conditions under which we obtained the
THz data confirm the excellence of the site, with the water vapor columns
ranging from 0.15 to 0.25~mm, resulting in a median single-sideband system
temperature of 1280~K. The telescope pointing was conducted on the strong
dust continuum of the background sources. The data were taken in wobbler-switching mode.
The FFT spectrometers delivered data with channel spacings of 76.3~kHz, smoothed to a velocity resolution of 1.7~\kms.
\subsection{Data reduction}
The data were calibrated with standard methods: Calibration loads at ambient and
liquid nitrogen temperature were measured approximately every 10 minutes to derive the count-rate to Kelvin conversion, allowing us to measure
the brightness temperature of the sky\footnote{Throughout this paper, all
temperatures are Rayleigh-Jeans equivalent temperatures, unless stated
otherwise.}; the atmosphere model reproducing the latter was then used for the
opacity correction. Details are described in \citet{2012A&A...542L...4G} and
\cite{2012MPI...MAN...12} for the calibration of data from GREAT and from the APEX,
respectively. For the aims of this work, that is, for absorption spectroscopy in front
of a continuum background source, the degree of accuracy of the absolute
calibration scale would be inconsequential for a single-sideband receiving
system. However, because of the double-sideband reception, an adequate correction
for the different atmospheric transmission in both sidebands is needed.
For the $\OHP$ lines, which are observed
in the upper sideband, the ratio of the signal-to-image band transmission is
close to unity for a sightline to the zenith under the weather
conditions mentioned above, and for our observations it has
a median value of 0.4. Thanks to the good observing conditions,
the ratio between the transmissions in the signal- to image-band were close to unity for GREAT.

The effect of the continuum calibration on the deduced column densities is discussed in the next paragraph. We provide a comparison with the continuum
calibration of PACS in Appendix \ref{app:F}.
   \begin{figure}[h!]
   \centering
   \includegraphics[width=0.85\columnwidth]{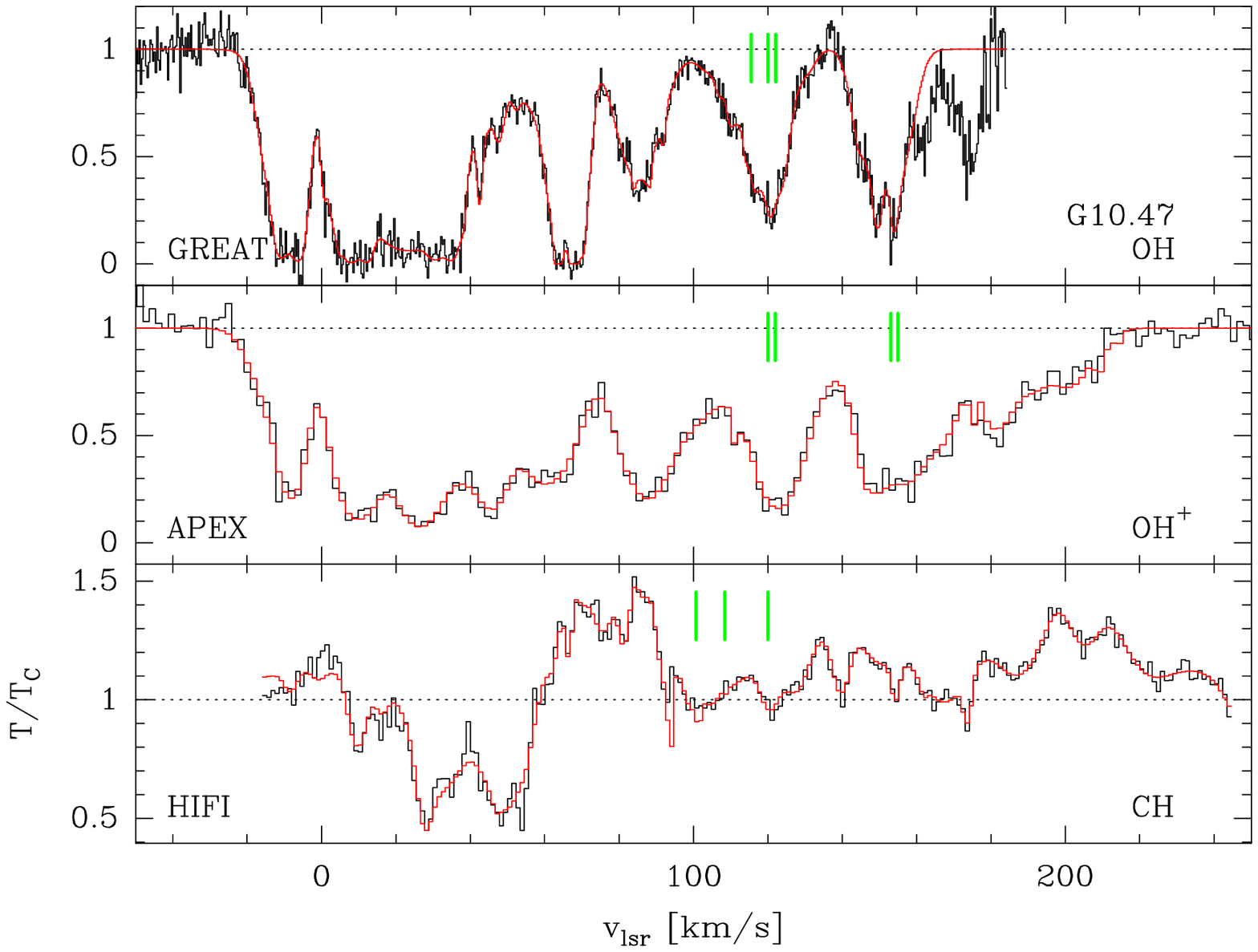}\vspace{2mm}
   \includegraphics[width=0.85\columnwidth]{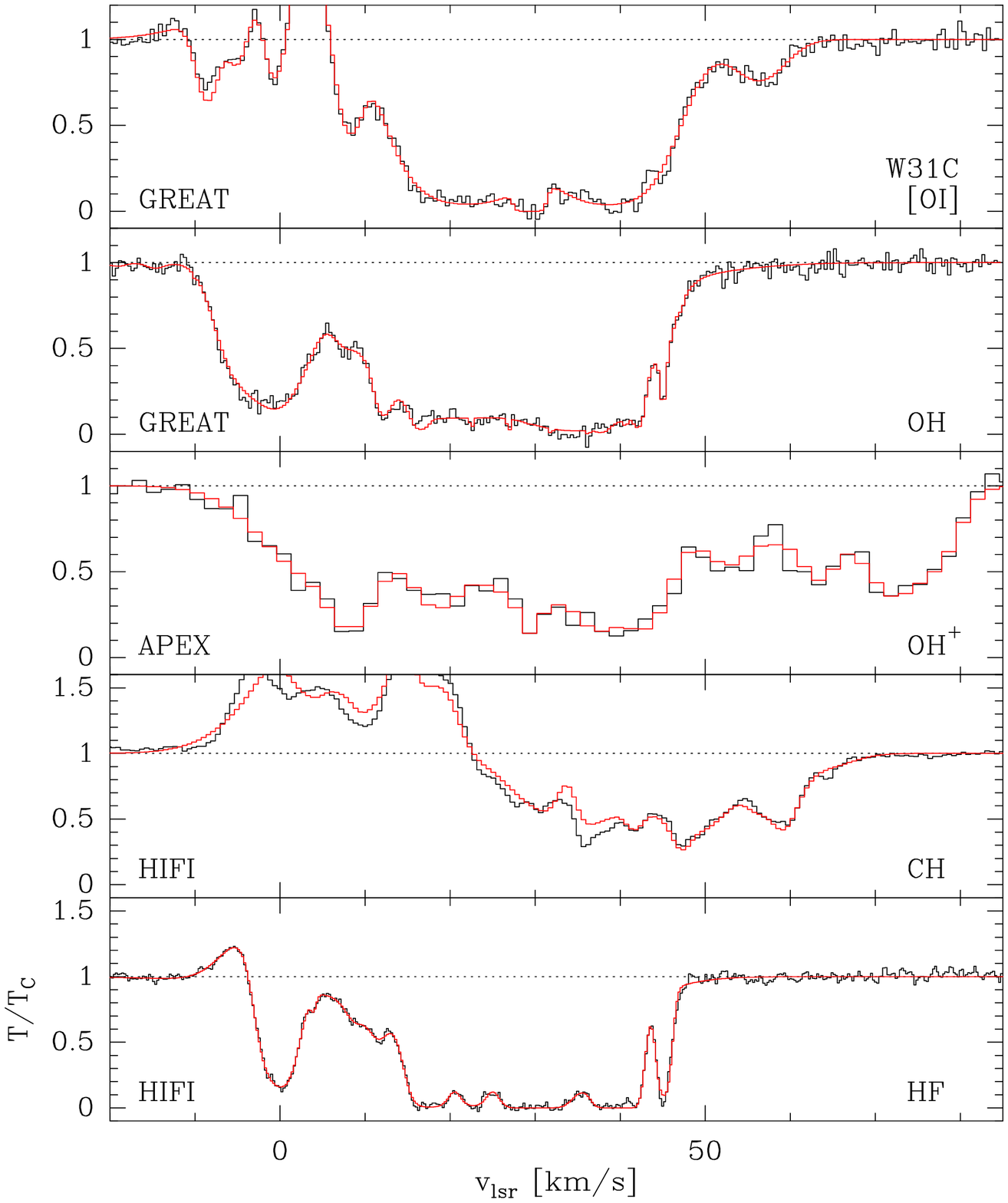}\vspace{2mm}
   \caption{\OI $^3P_1-^3P_2$, OH $^2\Pi_{3/2}\,\, J=5/2\leftarrow 3/2$
            and $\OHP$ $N=1\leftarrow0$ spectra of sightlines in the first
            quadrant, along with CH $^2\Pi_{3/2}\,\, J=3/2\leftarrow 1/2$
            and HF $J=1\leftarrow 0$ spectra. All five transitions
            are only available for W31C and W49N. Model fits are overlaid
            in red. For \OI, CH and HF one emission line
            component was added where needed. For G10.47, the hyperfine
            splitting is indicated at the 120~\kms component (green bars).
            All spectra are normalized by the continuum level.}
   \label{fig:spectra1stqu}
   \end{figure}
   \begin{figure}[h!]
   \addtocounter{figure}{-1}
   \centering
   \includegraphics[width=0.8\columnwidth]{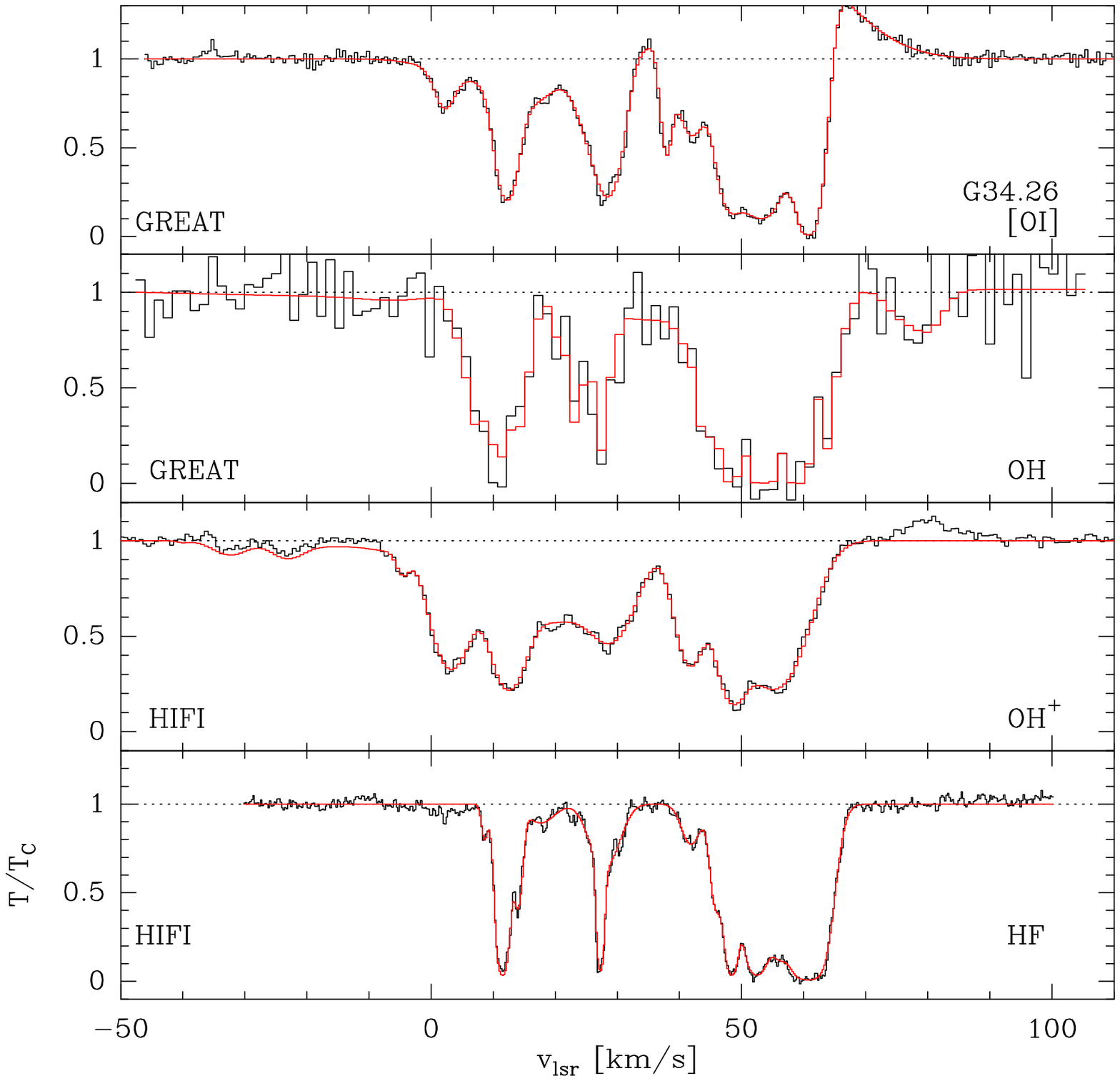}\vspace{2mm}
   \includegraphics[width=0.8\columnwidth]{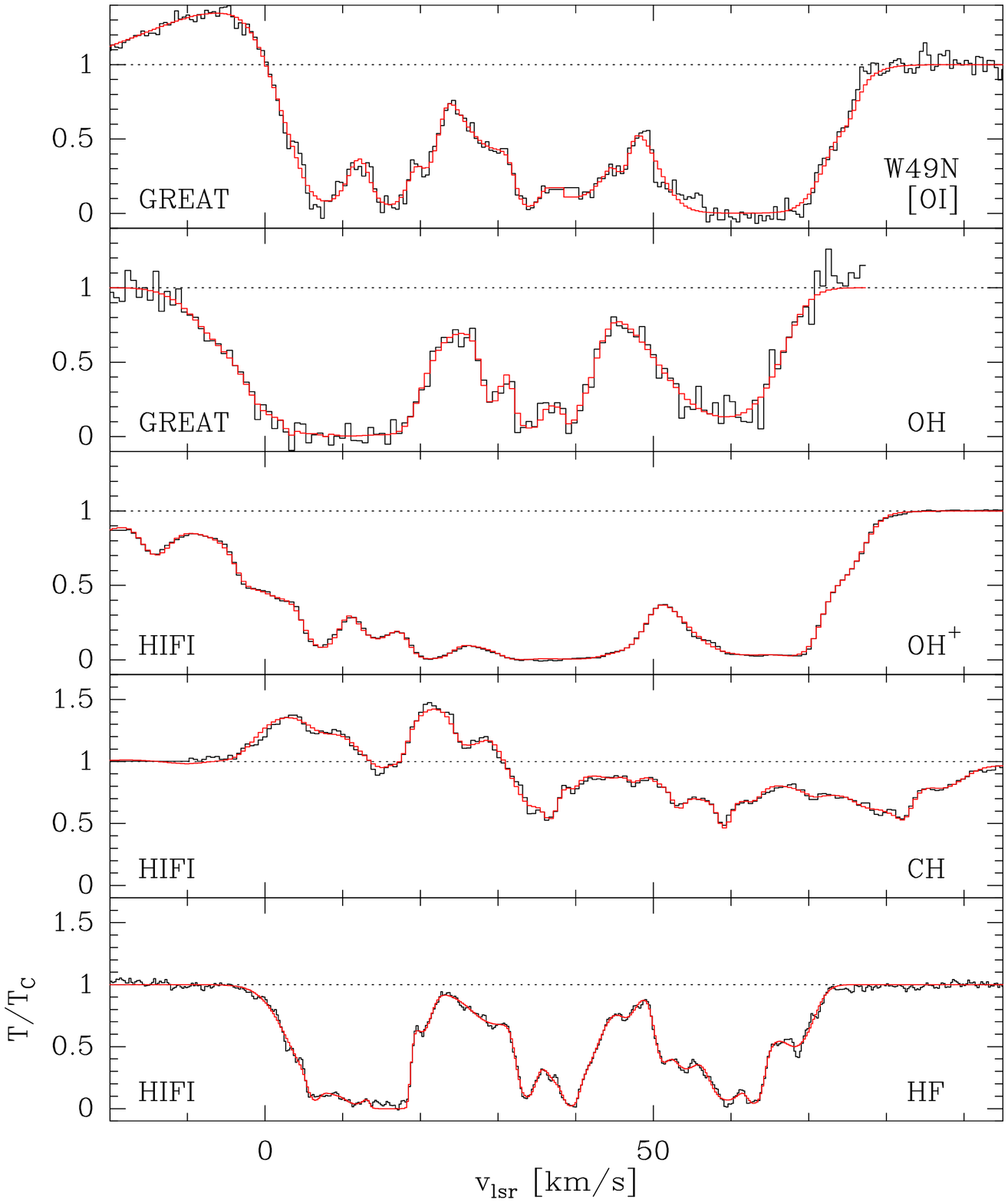}\vspace{2mm}
   \caption{ continued.}
   \end{figure}
   \begin{figure}
   \addtocounter{figure}{-1}
   \centering
   \includegraphics[width=0.8\columnwidth]{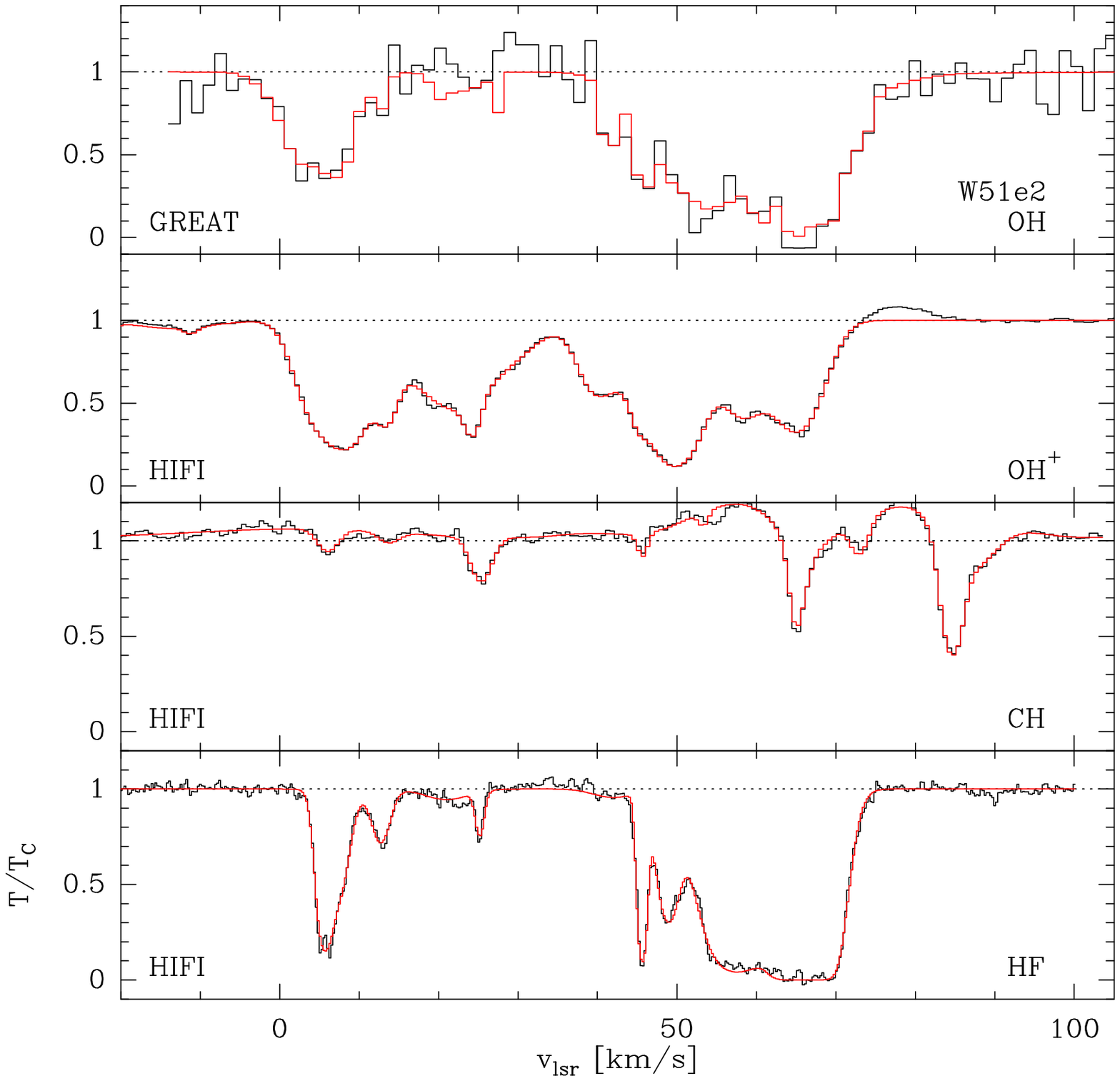}\vspace{2mm}
   \caption{ continued.}
   \end{figure}
   \begin{figure}[h!]
   \centering
   \includegraphics[width=0.8\columnwidth]{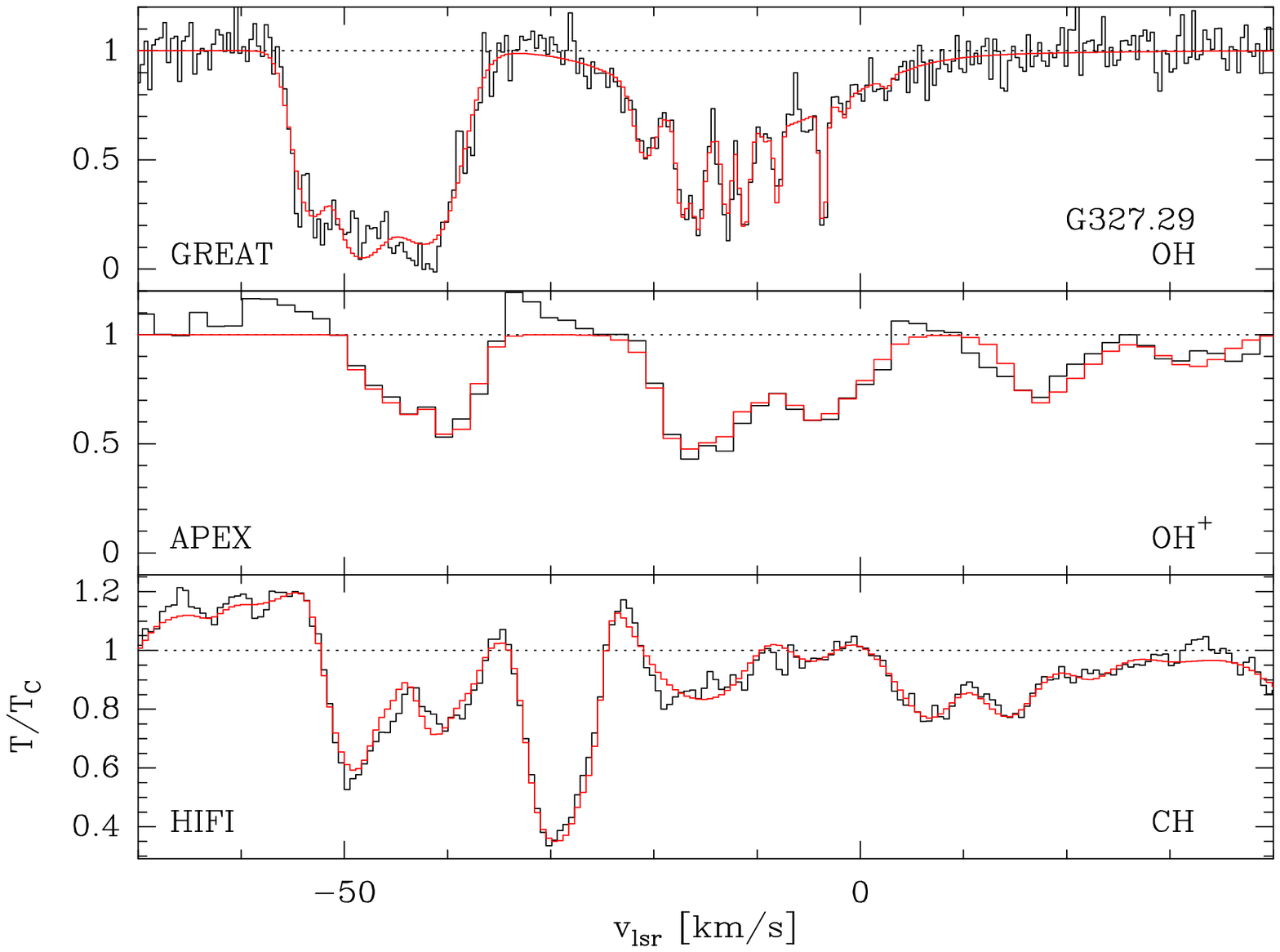}\vspace{2mm}
   \includegraphics[width=0.8\columnwidth]{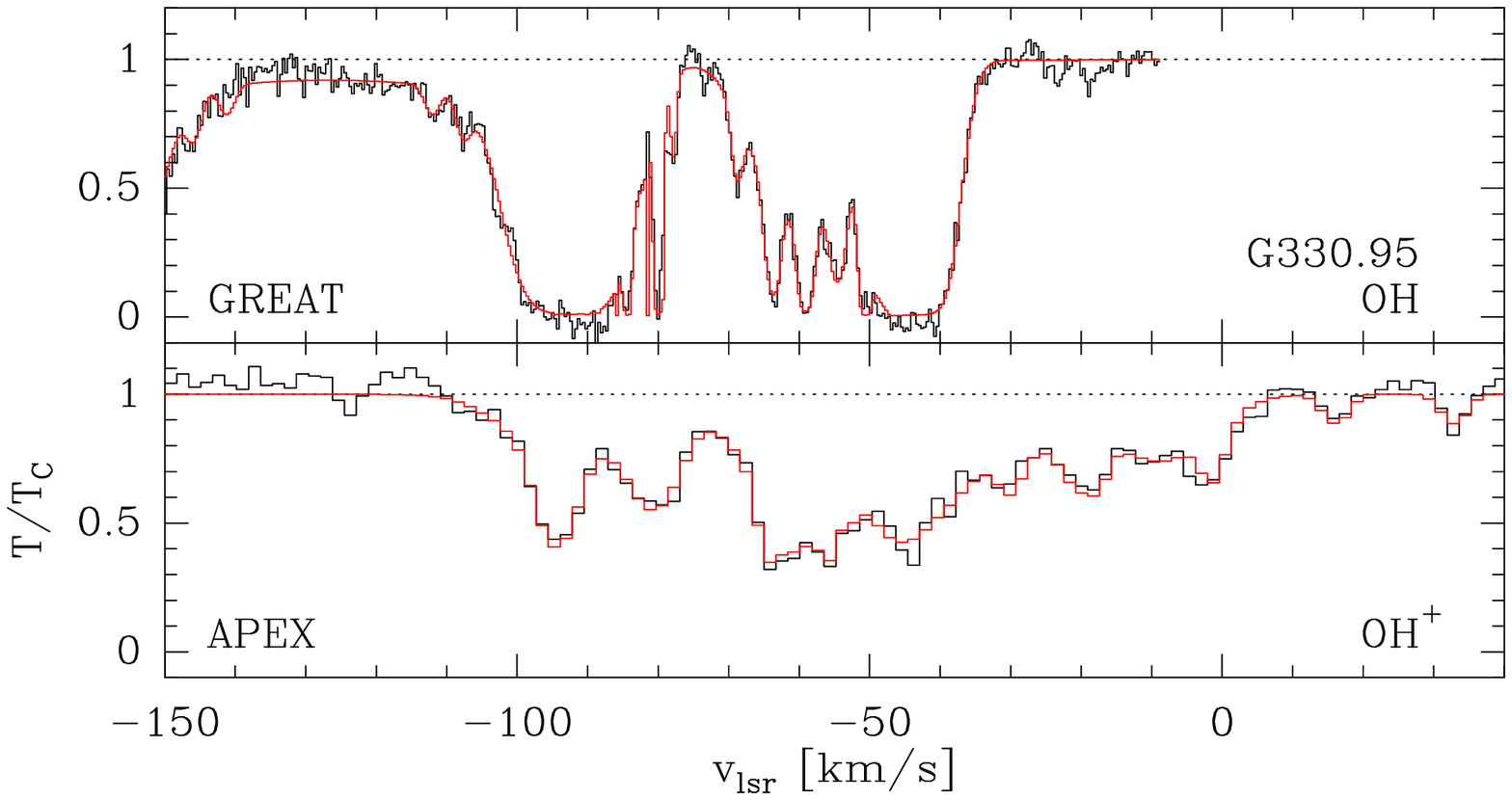}\vspace{2mm}
   \includegraphics[width=0.8\columnwidth]{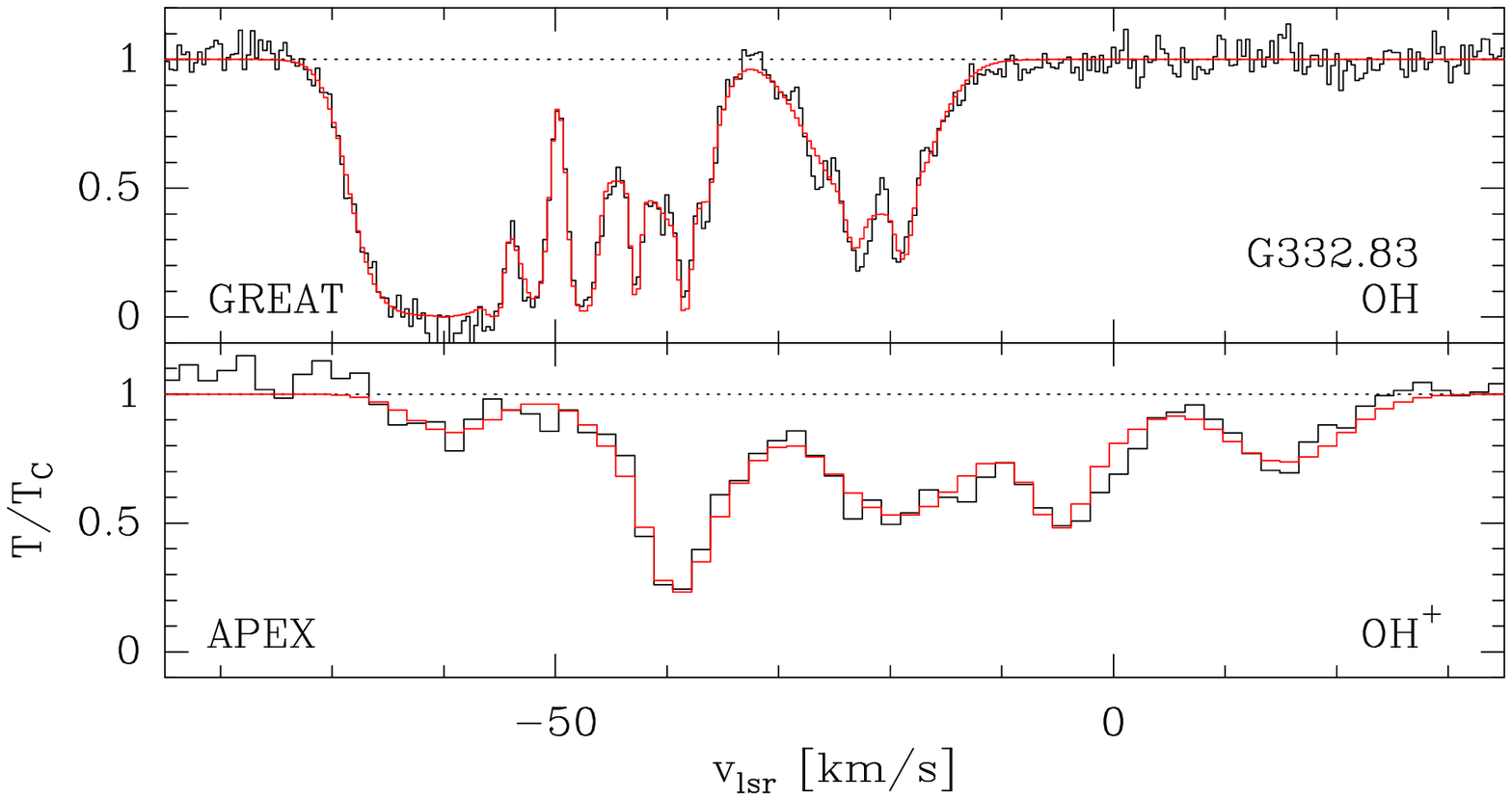}\vspace{2mm}
   \includegraphics[width=0.8\columnwidth]{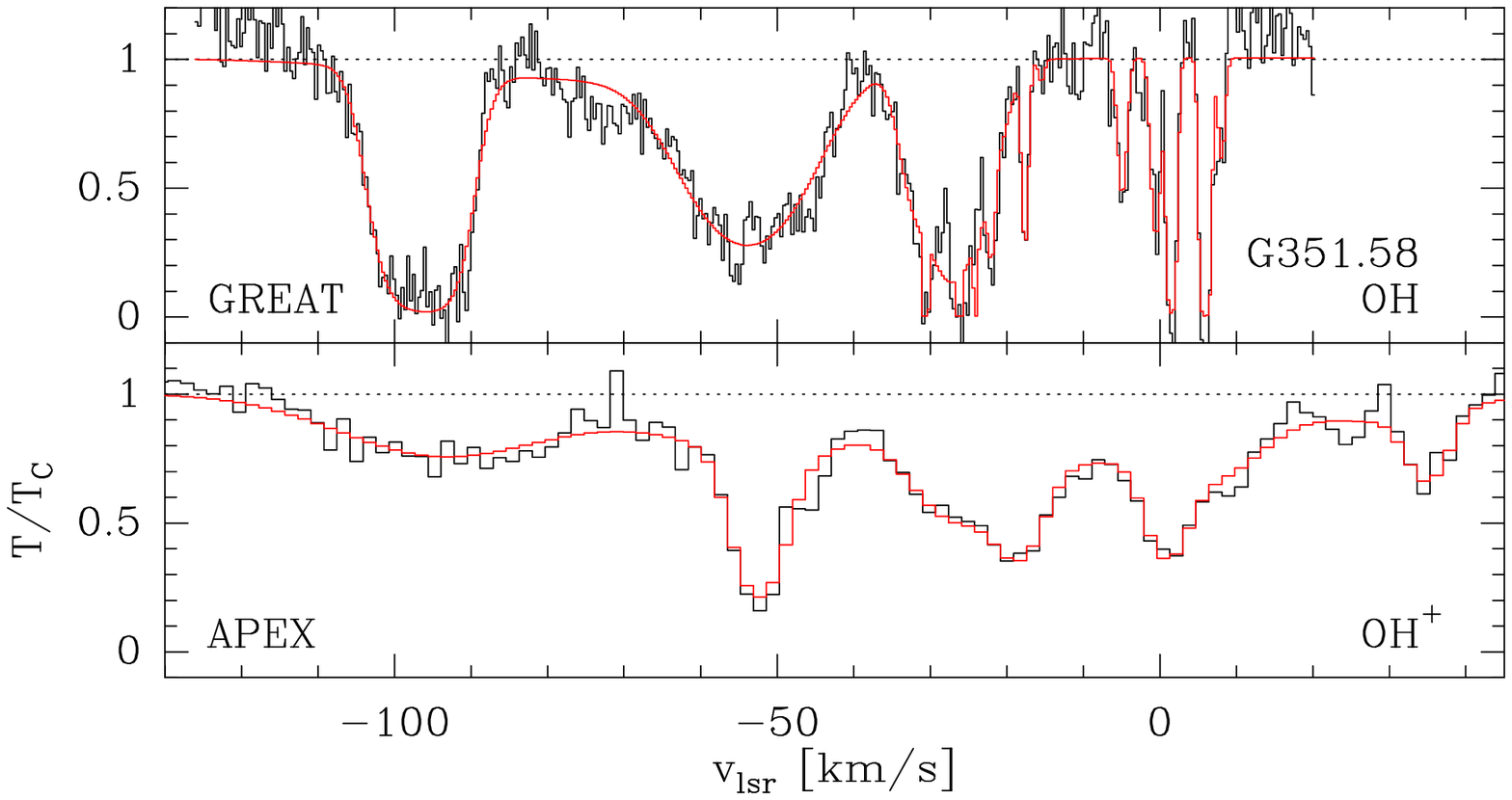}\vspace{2mm}
   \caption{Same as Fig.~\ref{fig:spectra1stqu}, but for sightlines in 
            the fourth quadrant. In G330.95, the OH absorption at
            velocities below $-130$~\kms is from the $\Lambda$ doublet
            in the other sideband and could be consistently fitted.}
   \label{fig:spectra4thqu}
   \end{figure}
   \begin{figure}[h!]
   \centering 
   \includegraphics[width=\columnwidth]{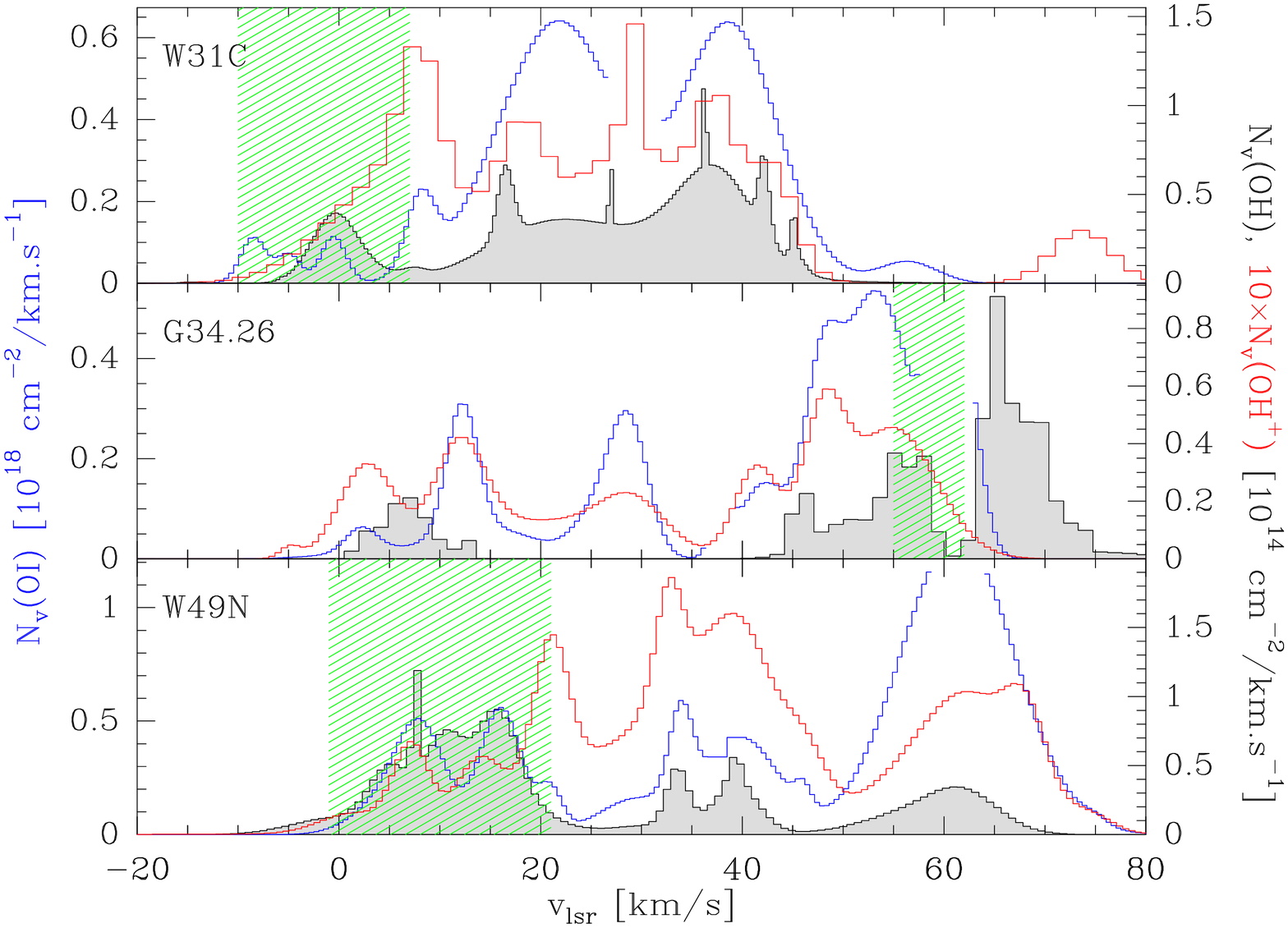}\vspace{2mm}
   \caption{Comparison of column densities of \OI (blue), OH (black and
            gray filled) and $\OHP$ (red, scaled by a factor 10) for sightlines to
            continuum sources in the first quadrant. The velocity range
            spanned by OH and CH$_3$OH masers in the hot cores is indicated
            by the green-hashed area. Velocity channels coinciding with the
            telluric or saturated \OI absorption are masked.}
   \label{fig:colden_o_oh_ohplus}
   \end{figure}
\subsection{Data analysis}
The power of ground-state absorption spectroscopy, namely, the direct
determination of column densities free from uncertain assumptions, has
been demonstrated in a number of publications (for OH, 
\citealp{2012A&A...542L...7W}). For a Gaussian line profile, the opacity of an
absorption line with several velocity and hyperfine components can be decomposed
and reads, for a velocity component $i$ and a hyperfine component $j$,
\begin{equation}
\tau_\mathrm{ij, \upsilon} =
\sqrt{\frac{\ln{2}}{\pi}} \frac{A_{\mathrm E,j}c^3}{4\pi\Delta \upsilon_{\rm i}
\nu_{\rm j}^3}
\frac{g_{\rm u,j}}{g_{\rm l,j}} N w_{\rm j}
\exp{
     \left(-4\ln{2}\left(\frac{\upsilon-\upsilon_{0,ij}}
                              {\Delta \upsilon_i}\right)^2\right)
    }
,\end{equation}
with the following quantities: 
the Einstein coefficient of the transition, $A_{\mathrm E,j}$, the full width
at half maximum of the Gaussian line profile, $\Delta \upsilon_{\rm i}$, the
degeneracies of the upper and lower levels involved in the transition,
$g_{\rm u,j}$ and $g_{\rm l,j}$, the total column density of the absorbing
molecular or atomic species, $N$, and the center velocity of the velocity
component, $\upsilon_{0,ij}$. The velocity scale of the obtained spectra
typically refers to the rest frequency of the strongest hyperfine component in
the local standard of rest; therefore $\upsilon_{0,ij}$ depends on the 
hyperfine component (the corresponding velocity offsets are listed in 
Table~\ref{table:1}). The factor $w_{\rm j}$ accounts for the fraction of species
that are in the corresponding ground state. For the physical conditions
prevailing in diffuse (atomic and molecular) and translucent clouds, that is, for gas
temperatures ranging from 15 to 100~K, densities from $\sim$10 to
$\sim$10$^3$~cm$^{-3}$ and extinctions from $A_{\rm V} = 0 - 2$
\citep{2006ARA&A..44..367S}, the ground-state occupation can be approximated by 
thermal equilibrium at the temperature $T_{\rm cmb}$ of the cosmic microwave
background (CMB), that is,
\begin{equation}
w_{\rm j} = g_{\rm j}/Q(T_{\rm cmb})
,\end{equation}
where $Q$ is the partition function. We assess the validity of this assumption
in Sect. 2.5 and conclude this paragraph with numerical details regarding our 
ansatz for the modeling, and the error analysis.

As a result of the blend of velocity components and hyperfine components, the data analysis is not straightforward.
The hyperfine structure can either be deconvolved \citep{2012A&A...540A..87G} or direcly fitted to the data. Here
we used the second option because deconvolution algorithms tend to produce spurious features
unless the spectral baselines are excellent, which is not the case for some data analyzed
here. Owing to the large number of free parameters, fitting many velocity components,
accounting for hyperfine blends, is not an easy task either, however. The XCLASS code
(https://www.astro.uni-koeln.de/projects/schilke/XCLASS), originally created for 
analyzing line surveys \citep[e.g.,][]{2005ApJS..156..127C}, is widely used for this
purpose and is based on the Levenberg-Marquardt algorithm (e.g.,
\citealp{1992nrca.book.....P}). An alternative method, used here, is the extension of 
discrete minimization by simulated annealing \citep{1953JChPh..21.1087M} to continuous,
nonlinear minimization problems with a large number of free parameters. The basic idea
\citep{1992nrca.book.....P} is to combine the Metropolis algorithm with the downhill
simplex method. To
avoid an undesired convergence to a local minimum, uphill steps are sometimes accepted,
with a probability defined by a "temperature" that is gradually lowered, in analogy to
the thermodynamic description of annealing. Both approaches have their strengths and
weaknesses. While their extensive discussion is beyond the scope of this paper, we conclude
that simulated annealing leads to satisfactory results once an efficient annealing scheme
(i.e., variation of the temperature along a sequence of iterations) is found. We also 
fit line profiles for species that do not display hyperfine structure
(HF, oxygen), for two reasons: First, we obtain a description of the underlying velocity
profiles, which are otherwise difficult to identify because the components blend with each other, and, second, the
granularity in the obtained column density profiles is avoided, while the $\chi^2$ function
of the fit allows estimating the errors.

The main-beam brightness temperature (Rayleigh-Jeans scale) of the 
observed spectrum is
\begin{equation}
T_{\rm RJ}(\upsilon) = T^{\rm (img)}_{\rm RJ, c} + T^{\rm (sig)}_{\rm RJ, c}
\exp{\left ( -\sum_{i=1}^{N_{\rm vc}}\sum_{j=1}^{N_{\rm hfc}} \tau_\mathrm{ ij, \upsilon} \right) }
\label{eq:radtran}
,\end{equation}
where $T^{\rm (sig)}_{\rm RJ, c}$ and $T^{\rm (img)}_{\rm RJ, c}$ are the 
Rayleigh-Jeans temperatures of the continuum in the signal and image band,
and $N_{\rm vc}$ and $N_{\rm hfc}$ are the number of velocity
components and hyperfine components, respectively. Uncertainties in the determination of the
signal band calibration temperature cancel out because they equally affect the
line and continuum temperature. The continuum level may also
be affected
(1) by the accuracy of the image band calibration and (2) by
standing waves whose power adds to that of the continuum. As mentioned above, case (1) could arise
from a slope between the atmospheric transmission in the signal and image band.
It can easily be shown that the resulting uncertainty in the opacity of the absorption line is (all temperatures and noise levels
are on the Rayleigh-Jeans scale, the subscript RJ is therefore omitted)
\begin{equation} 
\sigma_\tau = \sqrt{\sigma_{\rm T_c}^2+\left[\sigma_{\rm r}
T_{\rm c}^{(sig)}\right]^2} \frac{e^\tau}{T_{\rm c}^{\rm (sig)}}
,\end{equation}
where $\sigma_{\rm r}^2$ is the variance of ratio of the image band to the signal band
continuum. For the purpose of a rough error estimate we set
$\sigma_{\rm r} = 0.1$ as a conservative assumption, and
$T_{\rm c}^{\rm (sig)}e^{-\tau}$ to the observed line temperature $T_{\rm L}$
($=0$~K for saturated absorption, and $=T_{\rm c}^{(sig)}$ out of the line).
As expected, $\sigma_\tau \rightarrow \infty$ for a saturated line. As a bona fide estimate of
$\sigma^2_{\rm T_c}$ we can use the $\chi^2$ of the fit. Another approach, used 
in \citet{2012A&A...542L...7W}, consists of adding random noise to the fits to
the absorption line systems, and to determine the line parameters of these fits.
The distribution of fit parameters among a significant number of simulated
spectra yields empirical standard deviations. The advantage of this method is
that is does not require the errors to display a normal distribution. However,
for this study this approach is impractical because there are
so many
targets, sightline components, and species observed and analyzed. We therefore
used the error estimate outlined in this paragraph. In both approaches,
the errors are underestimated if instrumental bandpass ripples are
misinterpreted as weak but broad absorption features.

\subsection{OH level populations}
We finally examine our assumption that the level distribution can be described by
a thermal equilibrium with the CMB. We limit the discussion to OH, which is also an opportunity to
illustrate the difficulties encountered when radio lines are used for abundance
determinations. We apply two models (Table~\ref{table:2}) that are thought to bracket the conditions in diffuse
molecular and translucent clouds. The level population of OH was modeled with the
Molpop-CPE code written by \citet{2006MNRAS.365..779E}. Their idea is to extend the rate
equations of escape-probability methods with a sum of terms describing the radiative
coupling of the level populations between the various zones of the underlying sheet- or
slab-like geometries.
\begin{table}
\caption{Conditions in diffuse cloud models and coefficients for
departure from LTE of the population in OH $^2\Pi_{3/2}, J=3/2$ levels.
The FUV field is parametrized in units of the Habing field
\citep[Draine field = 1.7~Habing,][]{1978ApJS...36..595D}.}
\label{table:2}      
\centering          
\begin{tabular}{l c c}
\hline\hline       
                & Model 1           & Model 2        \\
\hline       
                & diffuse molecular  & translucent    \\
$\chi$ [Habing] & 1.7               & 1.7             \\
$A_{\rm V}$     & 0.2               &    1           \\
$n_{\rm H}$ [cm$^{-3}$]     & 100               & 1000           \\
$f^{\rm n}_{\rm H_2}$ & 0.1           &  0.5           \\
$T_{\rm gas}$ [K]  & 100               &   15           \\
$T_{\rm dust}$[K]  &  16               &   12           \\
\hline
 &\multicolumn{2}{c}{departure coefficients}  \\
\hline
$F=1-$ & 1.7580 & 0.9966 \\
$F=2-$ & 1.7565 & 0.9963 \\
$F=1+$ & 1.7398 & 0.9991 \\
$F=2+$ & 1.7384 & 0.9987 \\
\hline
\end{tabular}
\tablefoot{Defining quantities are from \citet{2006ARA&A..44..367S} and
from \citet{1999A&A...349L..25V} for the dust temperature. The clouds are 
immersed in the interstellar radiation field as given by
\citet{1983A&A...128..212M} and the cosmic microwave background.
Departure coefficients are defined as the fractional level population 
with respect to the population for thermalization at the CMB temperature, 2.73~K.}
\end{table}
   \begin{figure}[ht!]
   \centering
   \includegraphics[width=\columnwidth]{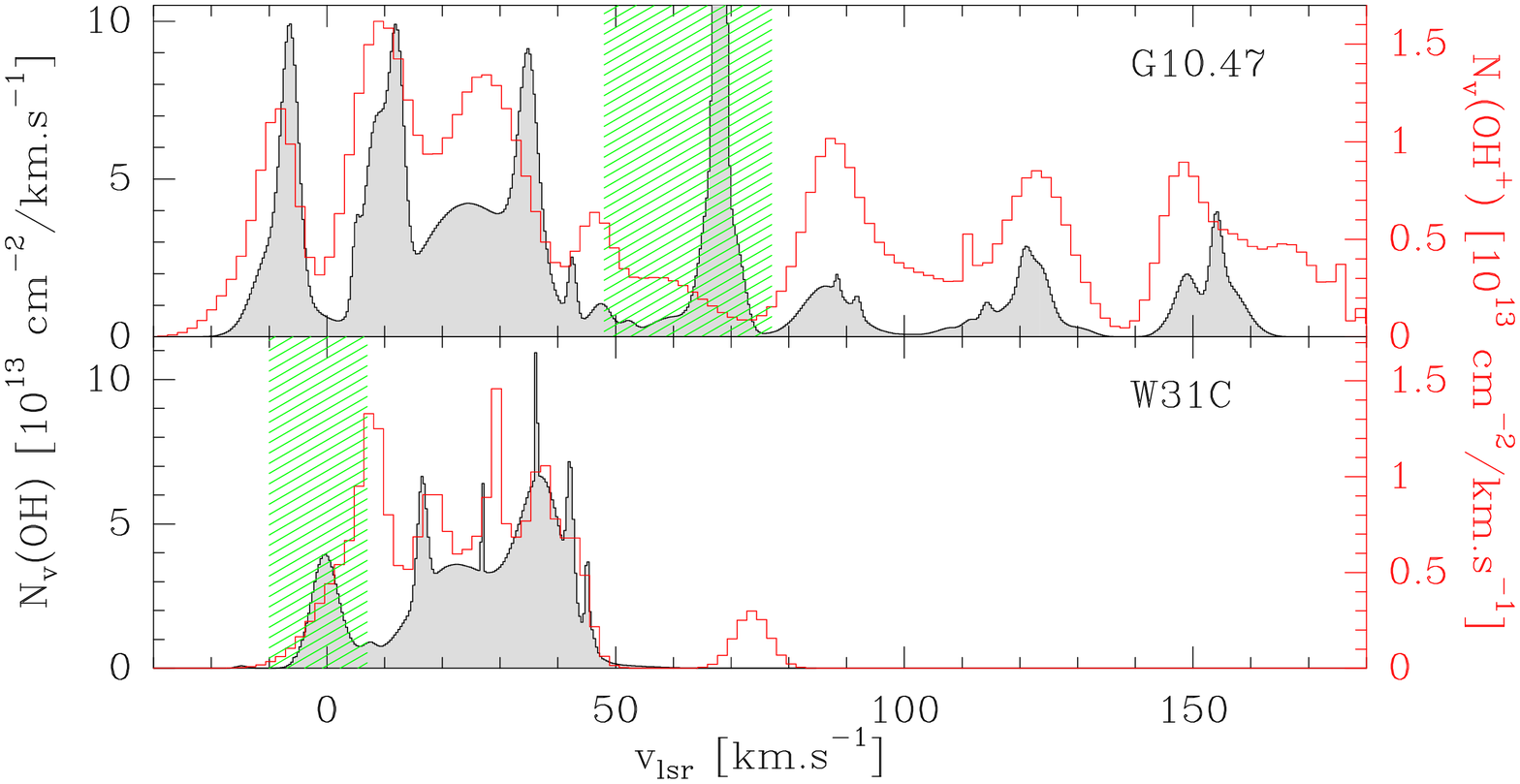}\vspace{2mm}
   \includegraphics[width=\columnwidth]{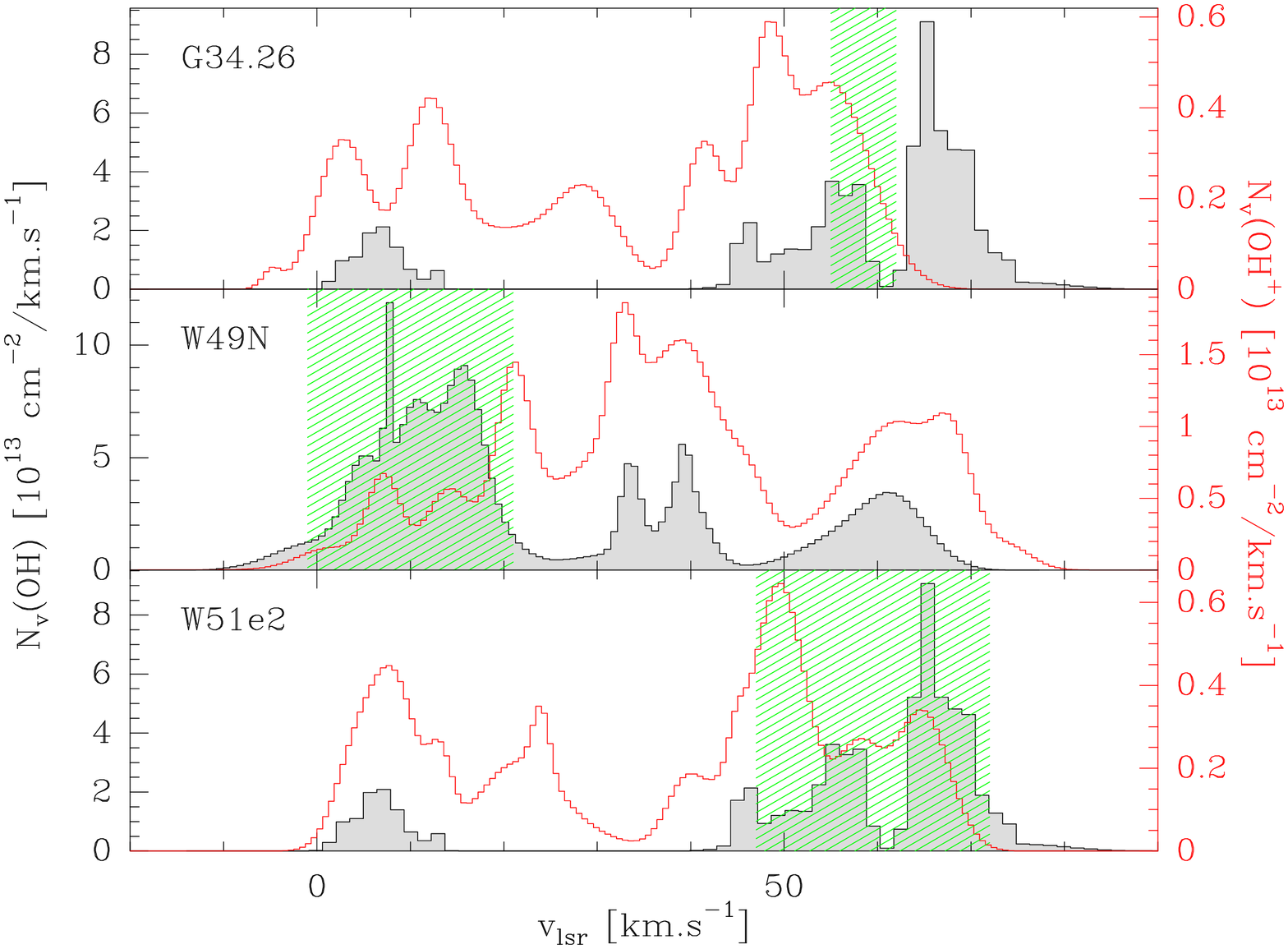}\vspace{2mm}
   \caption{Derived column density spectra of OH (gray filled histogram) and 
            $\OHP$ (red lines) toward continuum sources in the first
            quadrant. Here and in the following figures $N_{\rm V}$ denotes
            the velocity-specific column density, i.e., the column density within a
            velocity interval is the area under the histograms. The green hashed area indicates the velocity range
            spanned by the OH and CH$_3$OH masers in the hot cores where the derived column
            densities are uncertain due to saturated absorption and excitation
            of higher levels (for details see Table~\ref{table:3}).}
   \label{fig:colden1stqu}
   \end{figure}
\newpage
   \begin{figure}[h!]
   \centering 
   \includegraphics[width=\columnwidth]{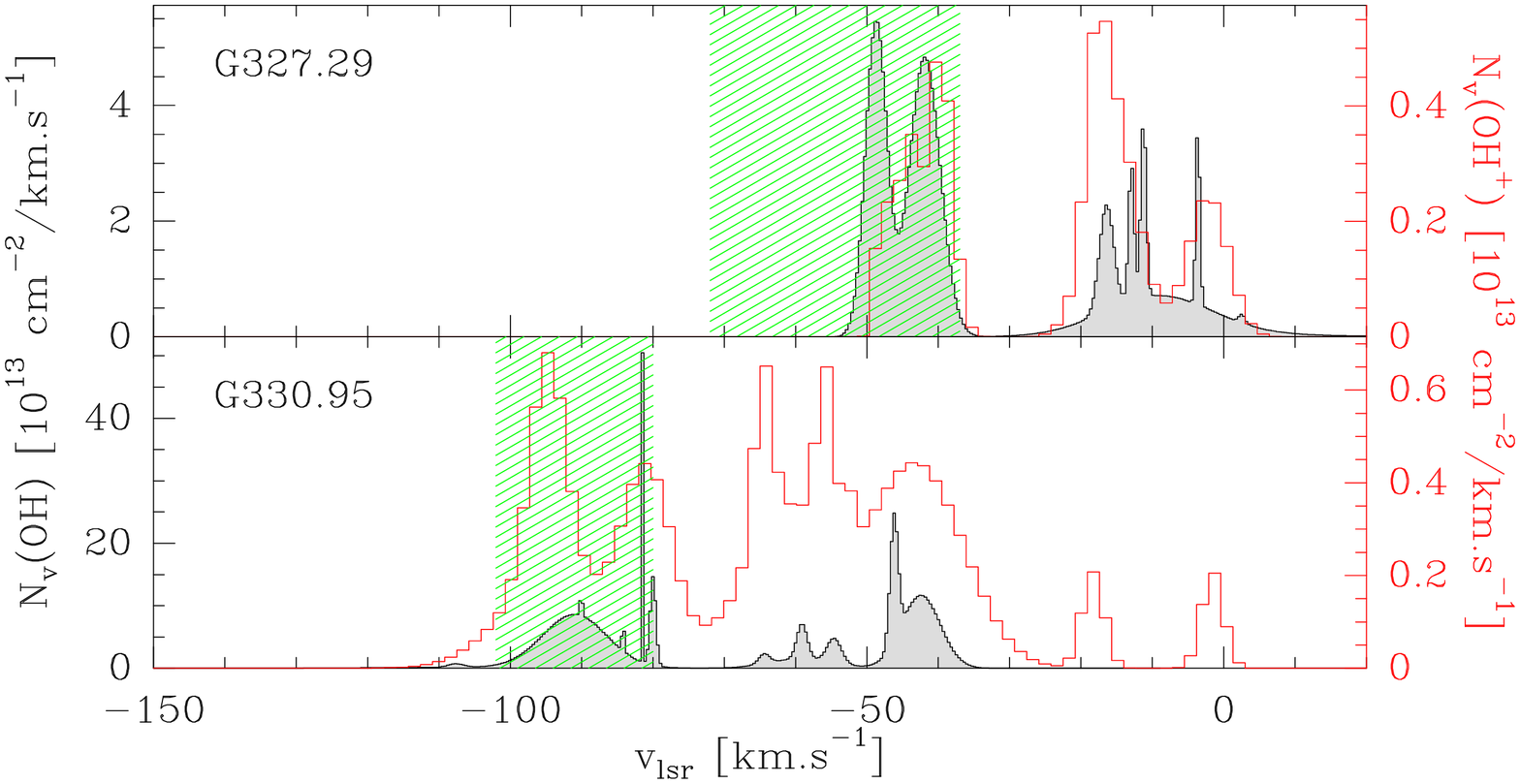}\vspace{2mm}
   \includegraphics[width=\columnwidth]{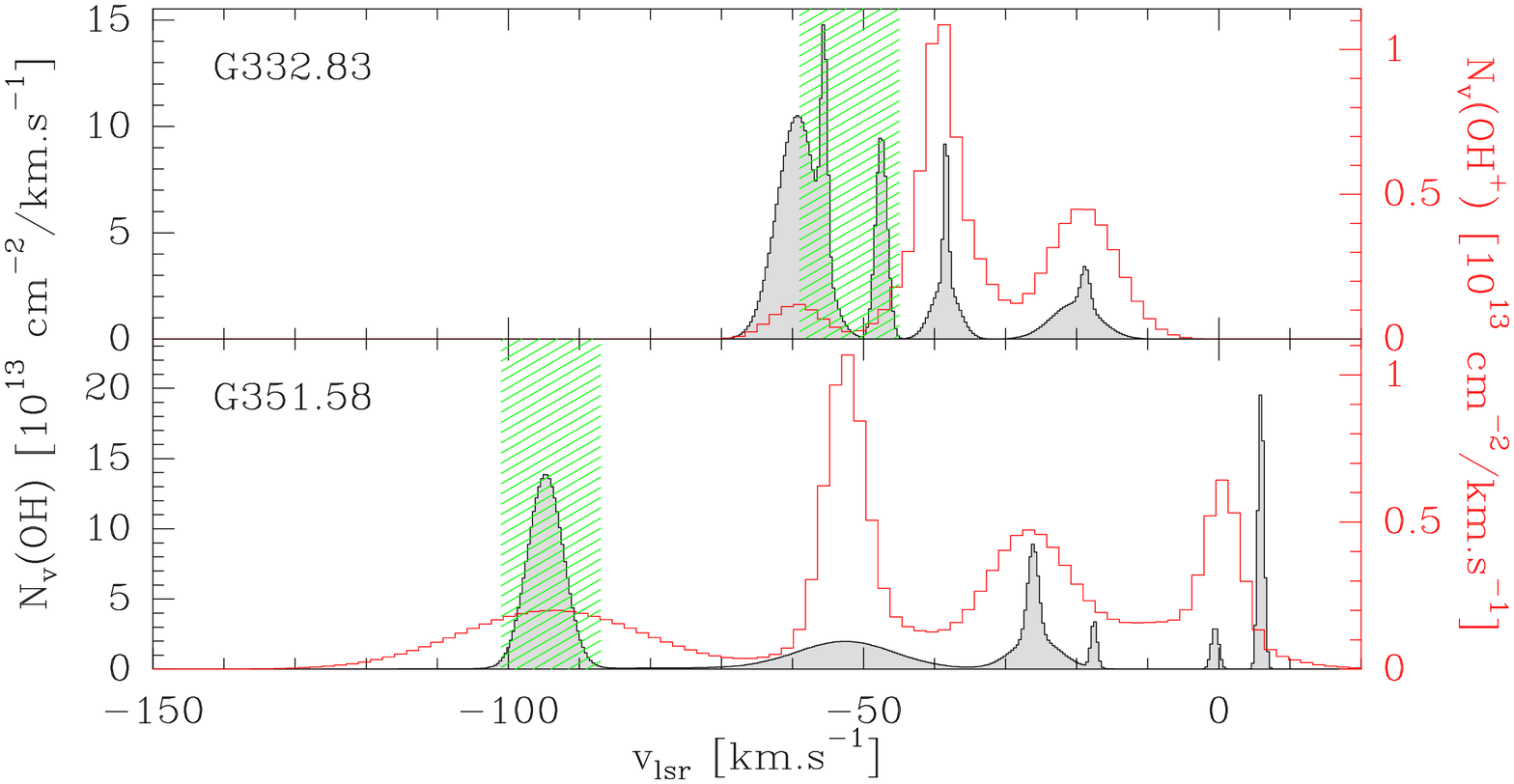}\vspace{2mm}
   \caption{As Fig.~\ref{fig:colden1stqu}, but for sightlines to
            continuum sources in the fourth quadrant.}
   \label{fig:colden4thqu}
   \end{figure}
The results show that while in model 2 the populations of
$^2\Pi_{3/2}, J=3/2$ states of OH can still be well described by a local
thermodynamic equilibrium (LTE) at the temperature of the cloud model, the 
assumption breaks down in model 1, with populations that are over-thermal
by a factor 1.7. This implies that excitation studies employing only the radio
lines of OH either describe the emitting gas inadequately or require NLTE modeling. 
On the other hand, in both models the departure of the level populations from LTE at
the temperature of the CMB (2.73~K) merely amounts to at most 1.024, resulting in a negligible error in the 
determination of OH column densities (we show below that uncertainties
in the calibration of the continuum level may lead to much larger errors). A
further study recalibrating the column densities of radio studies \citep[e.g.,][]{2002ApJ...580..278N}
with corrections deduced from far-infrared
observations is highly desirable, but is beyond the scope of this paper and will
be published separately.  In appendix \ref{app:D}, we present simple demonstrations
of the chemistry in these two cloud models.
\newpage
 \begin{figure*}
 \centering 
 \includegraphics[width=0.795\textwidth]{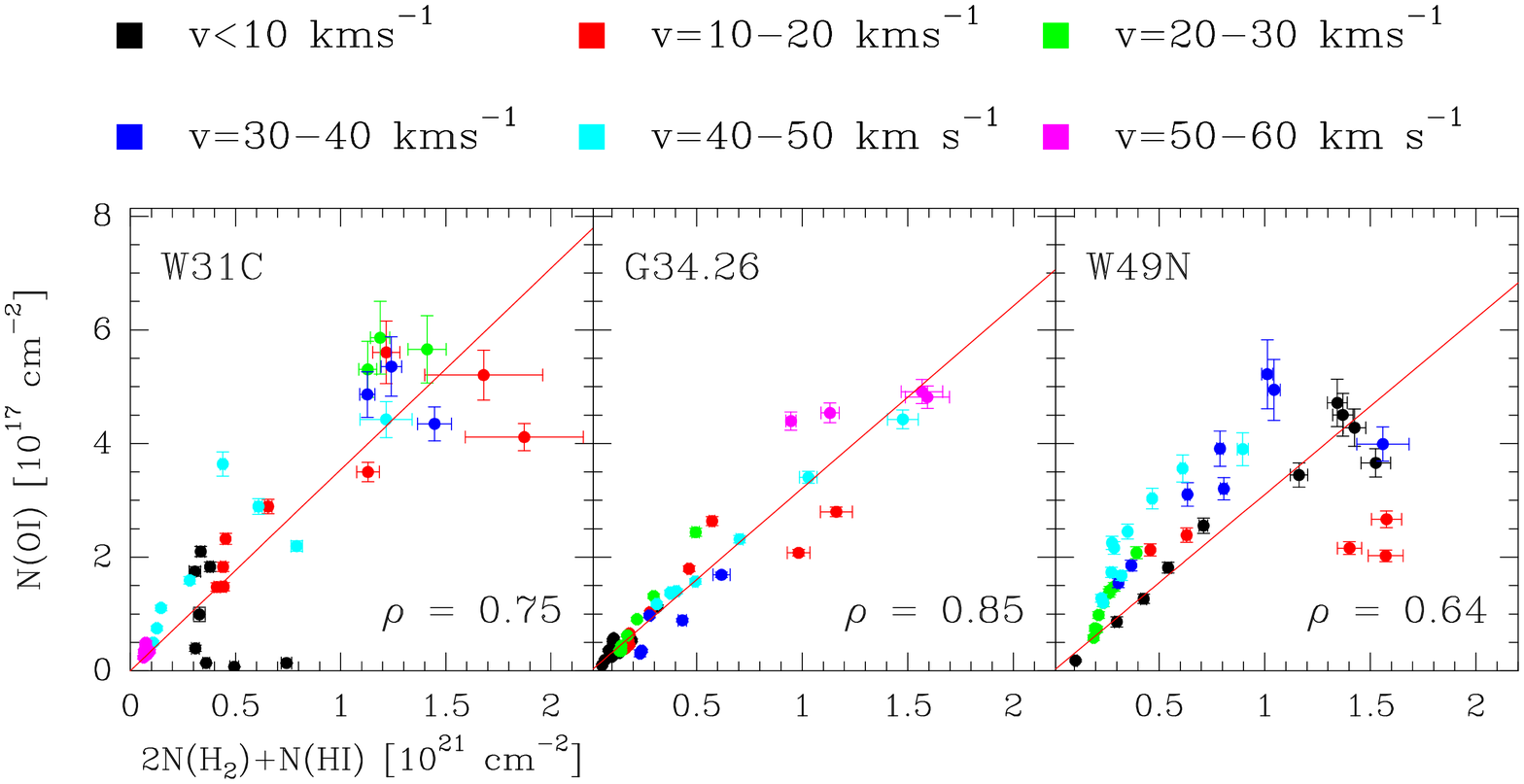}\vspace{2mm}
 \caption{Correlation between the \OI column density and the total
          (atomic and molecular) hydrogen column density derived from
          HF and \HI for sightlines toward W31C, G34.26, and W49N and in 
          1~\kms wide velocity bins. Colors indicate the velocity intervals
          defined in the legend. Only data points with $N/\sigma_{\rm N} > 5$
          both in \OI and $2N(\HH)+N$(\HI) are shown. The regression lines are shown in red.
          The corresponding correlation coefficients $\rho$
          are given in the lower right corners. For details see text.}
 \label{fig:oxygen_abundance}
 \includegraphics[width=0.795\textwidth]{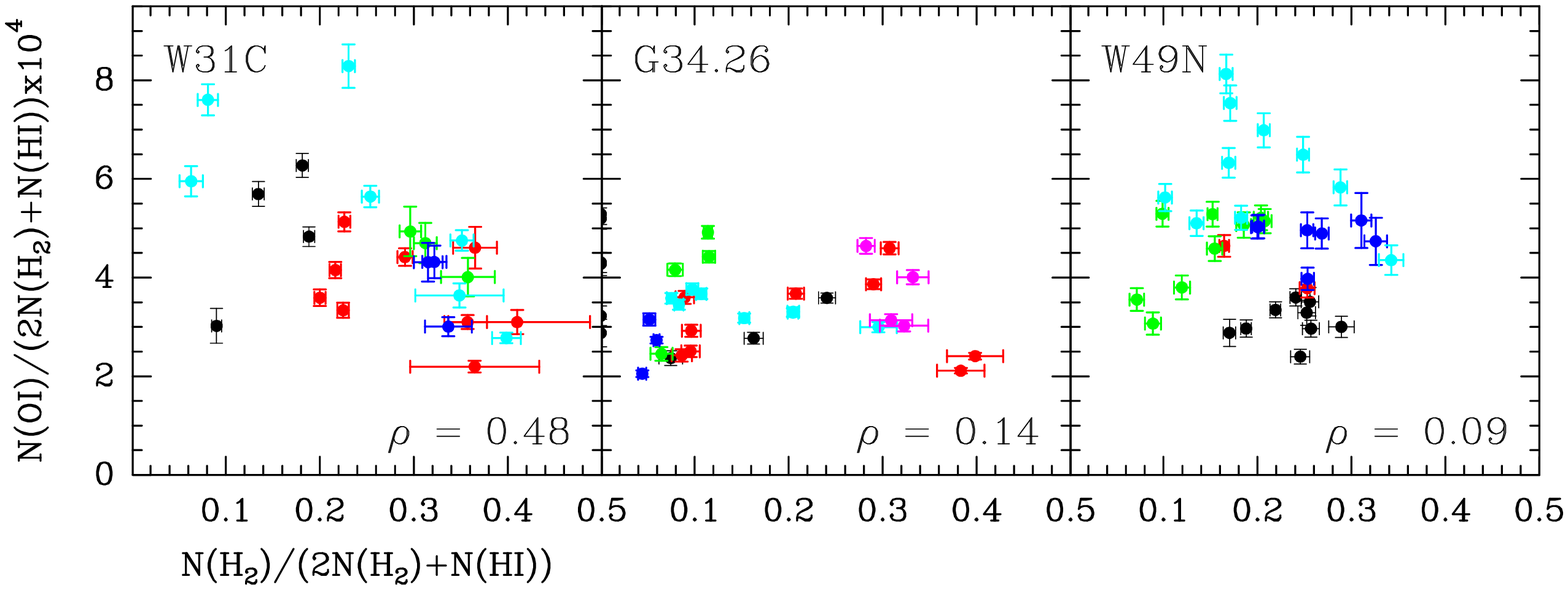}\vspace{2mm}
 \caption{Atomic oxygen abundance versus molecular hydrogen fraction $f^{\rm N}_\HH$,
          for sightlines toward W31C, G34.26, and W49N and in 
          1~\kms wide velocity bins. Correlation coefficients are given in the
          lower right corners. The correlation is significant only for W31C.
          Colors indicate the velocity intervals defined in the legend of
          Fig.~\ref{fig:oxygen_abundance}. Data points affected by saturated absorption
          are discarded.}
 \label{fig:oxygen_abundance_vs_fnh2}
 \end{figure*}
   \begin{figure*}
   \centering 
   \includegraphics[width=0.795\textwidth]{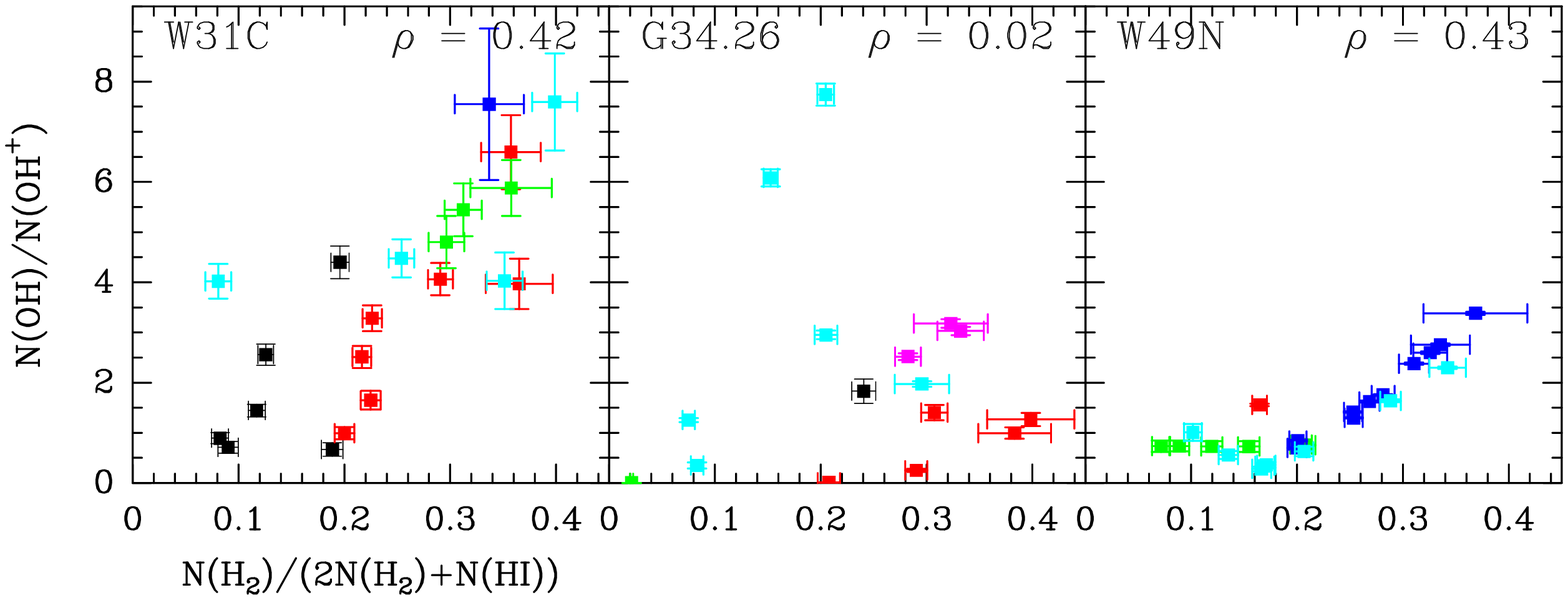}\vspace{2mm}
   \caption{Correlation between the $N(\OH)/N(\OHP)$ ratio and the molecular
            hydrogen fraction $f^{\rm N}_\HH = N(\HH)/(N($\HI$)+2N(\HH))$ for 
            W31C, G34.26, and W49N. Colors indicate the velocity intervals
            defined in the legend of Fig.~\ref{fig:oxygen_abundance}.
            Data points affected by saturated absorption are discarded.
            Correlation coefficients $\rho$ are given in the upper right
            corners. For details see text.}
   \label{fig:bottleneck}
   \end{figure*}
\section{Results}
\label{sec:data}
All spectra reveal complex absorption profiles, while \OI, HF, and CH are also 
seen in emission at the velocities of the hot cores and their environment, 
defined by OH and CH$_3$OH maser emission (Table~\ref{table:3}).
This also holds for the excited $^2\Pi_{1/2}, J = 3/2 \rightarrow 1/2$ line of OH,
often observed simultaneously with the OH spectra shown here \citep{2012A&A...542L...8C}. These
data will be published separately: The chemistry in the embedded ultra-compact HII regions is very different from
that of diffuse clouds and much more determined by endothermic reactions. 
Furthermore, the strength of absorption studies in ground-state transitions --
that is, column densities that can be directly inferred from observations --
breaks down in the high-excitation environment of hot cores.

Our results are presented and discussed in view of the velocimetry and
cartography of Galactic spiral arms, modeled by \citet{2008AJ....135.1301V}
in an effort to provide a symbiosis of available CO and \HI data. These
results were recently completed with the valuable distance measurements of
\citet[further references therein, see also below]{2014ApJ...783..130R}
that we adopt in the following. For the hot cores in the fourth quadrant, fewer parallaxes are known.
We use the distances of southern 6.7~GHz methanol masers obtained by \citet{2011MNRAS.417.2500G}
from Galactic kinematics, removing the distance ambiguity with \HI self-absorption, along with
the distances determined by \citet{2015arXiv150300007W}. We prefer sources in the 
Galactic plane, providing both strong continuum sources and allowing us to measure several 
spiral arm crossings in one spectrum, owing to the high resolution of GREAT.
The available spectra for sightlines in the first quadrant of the Galaxy are shown in Fig.~\ref{fig:spectra1stqu}
and the corresponding column density profiles in Fig.~\ref{fig:colden1stqu}.
The available spectra for sightlines in the forth quadrant of the Galaxy are
shown in Fig.~\ref{fig:spectra4thqu}, with the corresponding column density profiles
in Fig.~\ref{fig:colden4thqu}. In the following
$N_{\rm v}(Y)$ denotes the velocity-specific column density of species Y, that is, for a given velocity interval the
area below the histogram for species $Y$ is the column density. We refer to Appendix \ref{app:A} for
comments on the individual sightlines in the first and fourth quadrants.
\section{Discussion}
\label{sec:discussion}
For the following discussion we recall that all abundances and their ratios are derived from
column densities, not from volume densities of individual cloud entities. 
By consequence, it proves difficult to distinguish different clouds on the sightline.
Therefore our spectroscopy, although quantitative, cannot unambiguously identify
the various phases of the diffuse gas on the sightline. However, owing to the high spectral resolution of GREAT, we can separate
several spiral arm crossings, or identify interarm regions, by means of their characteristic velocity \citep{2008AJ....135.1301V}.
The deduced column densities and abundances within these regions are summarized in Tables~\ref{tab:results} and
\ref{tab:oxygen}. Abundances were determined from the average ratio of column densities across a given velocity interval, and
not from the ratio of averaged column densities. The rationale behind this is detailed in Appendix \ref{app:B}.
We use HF and CH as proxies for $\HH$. To assess the reliability of this approach, a correlation analysis
of $N(\HF)$ and $N(\CH)$ on the sightline to W49N was conducted (see Appendix \ref{app:E}) and yields a correlation coefficient
of 0.99, with a false-alarm probability below 1\%.

Column densities of \HI were obtained from JVLA data of the $\lambda$21~cm line (Winkel et al., in preparation).
\begin{table*}[ht!]
\caption{Synoptic summary of results. Column densities are integrated across the velocity intervals given in Col. 3.
Abundances are mean values across the indicated velocity intervals. Data affected by saturated absorption are not considered.}
\label{tab:results}
\begin{tabular}{l l c c c c c c c}
\hline\hline \\
sightline & spiral arm$^{(a)}$ & $\upsilon_{\rm min}, \upsilon_{\rm max}$ & $N(\OH)$ & $N(\OHP)$ & $N(\HH)\,^{(b)}$& X(OH)$^{(c)}$ & X($\OHP$)$^{(c)}$ & $\HH$ \\
          &            & [\kms]  &   \multicolumn{2}{c}{[$10^{14}$~cm$^{-2}$]} & [$10^{20}$~cm$^{-2}$] & \multicolumn{2}{c}{$\times 10^{-7}$} & proxy \\
\hline
G10.47    &            & $(-20,0)$   & $5.5\pm 1.8$    & $1.2\pm 0.1$     &   $10.7\pm 0.2$ &  $7.9 \pm 1.0$  &   $ 4.33   \pm 0.11$       & CH  \\
          & Sgr        & $(0,+50)$   & $19\pm 4$       & $4.8\pm 0.3$     &   $277\pm 3$    &  $0.83 \pm 0.17$    & $ 0.27 \pm 0.01$       & CH  \\
          & Gal. bar   & $(+80,+170)$& $7.3\pm 0.3$    & $4.6\pm 0.1$     &   $1167\pm 8$   &  $0.107 \pm 0.001$  & $  0.07\pm 0.01$     & CH  \\
G10.62$^{(d)}$ & Sgr   & $(+7,+15)$  & $0.95\pm 0.01$ & $0.67\pm 0.04$ & $8.21\pm 0.09$  & $1.25\pm 0.02$ & $1.12\pm 0.09$ & HF  \\
          & Sgr        & $(+42,+55)$ & $1.52\pm 0.05$ & $0.25\pm 0.02$ & $11.24\pm 0.26$ & $1.47\pm 0.13$ & $0.14\pm 0.01$ & CH  \\
          &            &             &                &                & $12.08\pm 0.34$ & $2.82\pm 0.79$ & $0.28\pm 0.02$ & HF  \\
          & Scutum     & $(+65,+80)$ & $-$    & $0.22\pm 0.01$  &         &              &     \\
G34.26    & Sgr        & $(0,+15)$   & $1.34\pm 0.15$ & $0.43\pm 0.02$   & $14.41\pm 0.47$         & $5.1 \pm 6.4$   & $0.69  \pm 0.59$ & HF  \\
          & Sgr        & $(+15,+30)$ & $< 0.001 $   & $0.26\pm 0.01$   & $ 9.39\pm 0.23$         & $< 0.005$     & $1.05  \pm 0.14$ & HF  \\
          &            & $(+38,+46)$ & $0.37\pm 0.01$ & $0.20\pm 0.01$   & $ 3.39\pm 0.07$         & $0.59 \pm 0.02$ & $1.35  \pm 0.54$ & HF  \\
          &            & $(+62,+85)$ & $5\pm 13$      & $0.023\pm 0.001$ & $13.19\pm 0.67$         & $49\pm 69$      & $0.048 \pm 0.027$         & HF  \\
W49N      & interarm   & $(+20,+30)$ & $0.83\pm 0.18$ & $0.93\pm 0.01$   & $22.9\pm 0.3$ &  $0.33 \pm 0.03$    & $ 0.40 \pm 0.01$      & CH  \\
          &            &             &                 &                 &   $4.2\pm0.1$   &  $2.22 \pm 0.20$    & $ 2.79 \pm 0.10$       & HF  \\
          & Sgr$^{(e)}$& $(+30,+45)$ & $3.9\pm 0.1$    & $2.17\pm 0.01$   &   $39.3\pm 0.9$ &  $0.93 \pm 0.03$    & $0.67  \pm 0.02$        & CH  \\
          &            &             &                 &                  &   $36.6\pm 0.7$ &  $1.06 \pm 0.02$    & $ 0.85 \pm 0.01  $      & HF  \\
          & Sgr$^{(f)}$& $(+45,+70)$ & $4.0\pm 0.1$    & $1.8\pm 0.1$   &     $70.3\pm 0.5$ &  $0.55 \pm 0.01$    & $ 0.42 \pm 0.01  $      & CH  \\
          &            &             &                 &                  &   $46.6\pm 0.5$ &  $1.54 \pm 0.09$    & $2.11  \pm 0.17$        & HF  \\
W51e2     & Sgr        & $(0,+15)$   & $1.4\pm 0.2$    & $0.43\pm 0.03$   &   $28.5\pm 7.0$ &  $0.49 \pm 0.11$    & $ 0.16 \pm 0.04$        & CH  \\
          &            &             &                 &                  &   $11.6\pm 0.1$ &  $4.41 \pm 4.47$    & $ 1.30 \pm 0.96$       & HF  \\
          &            & $(+15,+35)$ & $< 0.1$         & $0.28\pm 0.03$   &   $1.32\pm 0.07$&  $<1.0 $            & $2.92  \pm 0.72$        & HF  \\
          &            & $(+40,+50)$ & $0.91\pm 0.04$  & $0.37\pm 0.01$   &   $25.9\pm 5.1$ &  $0.28 \pm 0.06$    & $0.16  \pm 0.02$         & CH  \\
          &            &             &                 &                  &   $10.5\pm 0.2$ &  $1.17 \pm 0.15$    & $1.39  \pm 0.14$        & HF  \\
G327.29   & Carina     & $(-25,+10)$ & $2.6\pm 0.1$    & $0.58\pm 0.01$   &   $37.1\pm 19.8$&  $1.86 \pm 0.77$    & $0.77  \pm 0.43$        & CH  \\
G330.95   & Crux       & $(-70,-50)$ & $4.0\pm 0.3$    & $0.81\pm 0.02$   &   $89\pm 12$    &  $0.39 \pm 0.04$    & $0.09  \pm 0.01$        & CH  \\
          & Crux       & $(-50,-25)$ & $11\pm 1$       & $0.64\pm 0.01$   &   $84.4\pm 7.6$ &  $0.77 \pm 0.09$    & $0.16  \pm 0.01$       & CH  \\
          & Carina     & $(-25,+5)$  & $0.027\pm 0.005$& $0.08\pm 0.01$   &   $0.80\pm 0.19$&  $0.37 \pm 0.11$    & $0.97  \pm 0.84$        & CH  \\
G332.83   & Norma      & $(-74,-65)$ & $0.20\pm 0.11$    & $0.007\pm 0.001$ &                  &          &      &      \\
          & Crux       & $(-38,-30)$ & $0.55\pm 0.06$    & $0.31\pm 0.02$   &                  &          &      &      \\
          & Crux       & $(-30,-10)$ & $1.7\pm 0.1$    & $0.55\pm 0.01$     &           &          &      &      \\
G351.58   & Crux, Norma& $(-65,-35)$ & $3.3\pm 0.2$    & $1.0\pm 0.1$       &           &          &      &      \\
          & Carina     & $(-5,+5)$   & $0.42\pm 0.12$    & $0.42\pm 0.02$   &           &          &      &      \\
\hline
\end{tabular}
\tablefoot{$^{(a)}$ Spiral arms: Sgr/Car Sagittarius-Carina arm, Crux/Scu Crux-Scutum
arm; $^{(b)}$ Using $[\CH]/[\HH] = 3.5\times 10^{-8}$ \citep{2008ApJ...687.1075S} and $[\HF]/[\HH] = 1.4\times 10^{-8}$
(cf. Appendix~\ref{app:E}); $^{(c)}$ velocity-interval averaged abundances with respect to $\HH$; $^{\rm (d)}$ alias W31C
(the CH spectrum was not used for the $\upsilon_{\rm lsr} = (7,15)$~\kms interval due to blend with emission);
$^{\rm (e)}$ near- and far-side crossing of Sagittarius arm, $^{\rm (f)}$ far-side crossing.}
\end{table*}
\subsection{Distribution of \OI}
Spectroscopically resolved observations of the $^3P_1 \rightarrow ^3P_2$ fine structure line of atomic oxygen, \OI, at $\lambda$63.2~$\mu$m are rare.
After the pioneering work of \citet{1979ApJ...227L..29M}, who discovered the line in M17 and M42 with NASA's Lear Jet, \citet{1996ApJ...464L..83B} resolved
the line in M17 at 0.2~\kms channels spacing with laser heterodyne spectroscopy onboard the KAO. Such a high spectral resolution was out of reach
for the ISO-LWS observations, with $\sim$35~\kms. Nevertheless, \citet{2001ApJ...561..823L} successfully separated the \OI absorption toward Sgr B2 into four
components, three of which were found to be located in foreground clouds. By correcting the \OI abundance in the atomic gas phase with existing \HI data, the
authors were able to establish an \OI to CO abundance ratio of $\sim$9 , with a 50~\% error. The deduced \OI abundance relative to the total hydrogen reservoir (i.e.,
atomic and molecular) amounts to $2.7\times 10^{-4}$ in the dense molecular gas. Meanwhile, the spectroscopy of the ground-state lines of HF and of
$^2\Pi_{3/2}$~CH has become accessible.

From the \OI column densities shown in Fig.~\ref{fig:colden_o_oh_ohplus} and using
HF as a proxy for H$_2$, we obtain the correlation between $N$(\OI) and $N($\HI$)+2N({\rm H}_2)$ shown in Fig.~\ref{fig:oxygen_abundance}.
Data points with $N/\sigma_{\rm N} < 5$ in both quantities are discarded from the correlation analysis and the figure. This conservative
cutoff implies that our analysis is not affected by saturated absorption. 
The false-alarm probabilities are below 5\%; here and in the following correlation analyses they are assumed to be given by
Pearson's p-value. Ranging from 0 to 1, it provides an estimate of the probability that the two quantities are uncorrelated (which is
the null hypothesis, e.g., \citealp{1995inco.book.....C}). From a weighted regression analysis we derive oxygen abundances
(relative to the total hydrogen reservoir, atomic and molecular) of $X($\OI$) = (3.5\pm 0.1)$, $(3.2\pm 0.1)$ and $(3.1\pm 0.1)\times 10^{-4}$
for W31C, G34.26, and W49N, respectively. Velocities where the absorption in HF or \OI is saturated are excluded, as are those corresponding
to the hot cores, where the assumption of a complete ground-state population does not hold anymore. Remarkably, the abundances along these
very different sightlines agree within $2\sigma$. Likewise, \citet{1998ApJ...493..222M} determined from UV spectroscopy of the $\lambda 1356${\AA}
line toward 13 distinctly different sightlines tracing low-density diffuse gas $A_{\rm V} \la 1.0$ mag an oxygen abundance of
$(3.19\pm 0.14)\times 10^{-4}$, in agreement with our result. \citet{2004ApJ...613.1037C} later used the same technique on 36 sightlines and
found oxygen abundances of $3.9\times 10^{-4}$ on sightlines with lower mean density and $2.8\times 10^{-4}$ on the denser ones.
\citet{2005ApJ...619..891J} determined a value of $4.2\times 10^{-4}$ from their ten best-determined sightlines, with a
$\sim 10$\% error. For the sake of comparison, we recall that the solar value was corrected downward over the years from
$7\times 10^{-4}$ \citep[][further references therein]{1973ASSL...40....1P} to $4.9\times 10^{-4}$ \citep{2009ARA&A..47..481A}.

All oxygen abundances are encompassed at the lower end by the value of \citet{2001ApJ...561..823L} for the denser, mainly molecular gas (but they are still
compatible with our values within their error bars), and at the higher end by stellar abundances, for instance, $(5.75\pm 0.4)\times 10^{-4}$ in the B stars of
nearby OB associations \citep{2008ApJ...688L.103P}. The variation of abundances is partly due to inhomogeneities in the Galactic
distribution of oxygen and partly due to the different analysis techniques. Observations from near-infrared to UV wavelengths need to apply corrections
for interstellar extinction and non-LTE effects; neither correction is necessary in our approach using the $\lambda$63.2~$\mu$m \OI fine structure
line. Abundances derived from ionized gas phases need to account for the resonant charge transfer reaction O+(H,H+)O
\citep[e.g.,][]{2011piim.book.....D}, while our determination refers to the diffuse neutral gas phase, with an ionization fraction much smaller than unity 
(with $n_{\rm e}/n_{\rm H} \sim 10^{-4}$, mainly from the photoionization of carbon, e.g., \citealp{1998ISAA....4...53V}).
At the lower end of the determined oxygen abundances, that is, in clouds with a higher fraction of molecular hydrogen, the differences are
seemingly easier to understand: At $A_{\rm V} \ga 1.0$~mag, the \OI reservoir is used for the synthesis of oxygen-bearing species and
for inclusion in the ice mantles of dust grains (while only some oxygen is fed back to the gas, e.g., through photodissociation of OH).
However, our picture of oxygen depletion is incomplete. Adopting as reference oxygen abundance the value of
$(5.75\pm 0.4)\times 10^{-4}$ of \citet{2008ApJ...688L.103P}, \citet{2010ApJ...710.1009W} compared abundance measurements in various oxygen 
reservoirs to densities of up to $\sim$1000~cm$^{-3}$ and found that an additional source of oxygen depletion is required at densities of
$\ga$10\,cm$^{-3}$. At the transition between diffuse and molecular clouds, where the opacity of the ISM precludes UV spectroscopy and where ice
mantles start to form around dust grains, the author found that an oxygen abundance of $1.6\times 10^{-4}$ cannot be attributed to either silicates
or oxides. The carrier of this unidentified depleted oxygen (UDO) may be oxygen-bearing, organic matter. \citet{2009ApJ...700.1299J}
concluded the same in a UV study
of 17 elements on 243 sightlines. While this issue requires further investigation, we note that our oxygen abundance
of $3\times 10^{-4}$ precludes a strong increase of UDO at the densities sampled here (i.e., $n_{\rm H}\sim$100 to 1000~cm$^{-3}$), but does not rule out the presence of UDO either.

In summary, our abundance determination (on sightlines with molecular hydrogen fractions $f^{\rm n}_{\rm H_2}\la 0.4$, with most data points below the
value of 0.25 where half of the hydrogen atoms are bound in $\HH$) is compatible with the somewhat larger oxygen abundance in gas that is
predominantly atomic, in agreement with this picture. So far, only for W31C our data show an anti-correlation between the oxygen
abundance and the molecular hydrogen fraction $f^{\rm N}_\HH$ with a p-value $< 0.05$ (Fig.~\ref{fig:oxygen_abundance_vs_fnh2}; again, only data points
with a relative error <20\% are retained). This significance may be taken as an indication of the depletion of \OI with increasing molecular hydrogen
density. However, the corresponding correlations on the sightlines to G34.26 and to W49N are not significant. As a caveat, we point out that these
results depend on the adopted abundance of HF (which is the main surrogate for $\HH$) and therefore on the underlying assumption that it is uniform
along various sightlines. Therefore, these results are to be interpreted with caution. For G34.26 and W49N the correlations shown in
Fig.~\ref{fig:oxygen_abundance_vs_fnh2} are insignificant, with p-values of 0.42 an 0.58, respectively.
This suggests that \OI in the diffuse atomic and molecular gas is a tracer of both atomic and molecular hydrogen, except at the densities of
translucent clouds and above. The formation of molecular oxygen as a source of depletion of \OI
is negligible here: While O$_2$ has not been detected in the diffuse ISM, its abundance falls short of predictions by at
least two orders of magnitude \citep{2005AdSpR..36.1027M}. Finally, we stress again that the \OI abundances derived here are sightline-averaged
quantities. The scatter in Figs.~\ref{fig:oxygen_abundance} and \ref{fig:oxygen_abundance_vs_fnh2} might also be due to the galactocentric oxygen
abundance gradient that is a natural consequence of oxygen production in short-lived, massive stars. In our sightlines the corresponding variation
of the oxygen abundance may be of a factor of up to two: From stellar \OII spectra, \citep{1997ApJ...481L..47S} measure a gradient of $-0.07$~dex/kpc.
From their NLTE analysis of the NIR \OI triplet \citet{2014MNRAS.444.3301K} obtained a similar gradient of $-0.058$~dex/kpc, and found
\begin{equation}
\log{[\O/\H]} =  -2.72-0.058\,R_{\rm G}\,[\mathrm{kpc}]\,.
\label{eq:oxygen}
\end{equation}
For the solar $R_{\rm G} = 8.34$~kpc \citep{2014ApJ...783..130R}, the local oxygen abundance amounts to $X($\OI$) = 6.3\times 10^{-4}$. The
smallest galactocentric radius among our \OI sightlines is that of W31C, $R_{\rm G} = 3.6$~kpc, where
$X($\OI$) = 1.2\times 10^{-3}$ is expected. An attempt to determine the Galactic \OI gradient from our ground-state data remained
inconclusive because of the limited number of sightlines and the frequent saturated absorption, but should be possible with a dedicated, extended
study.
   \begin{figure}[ht!]
   \centering 
   \includegraphics[width=\columnwidth]{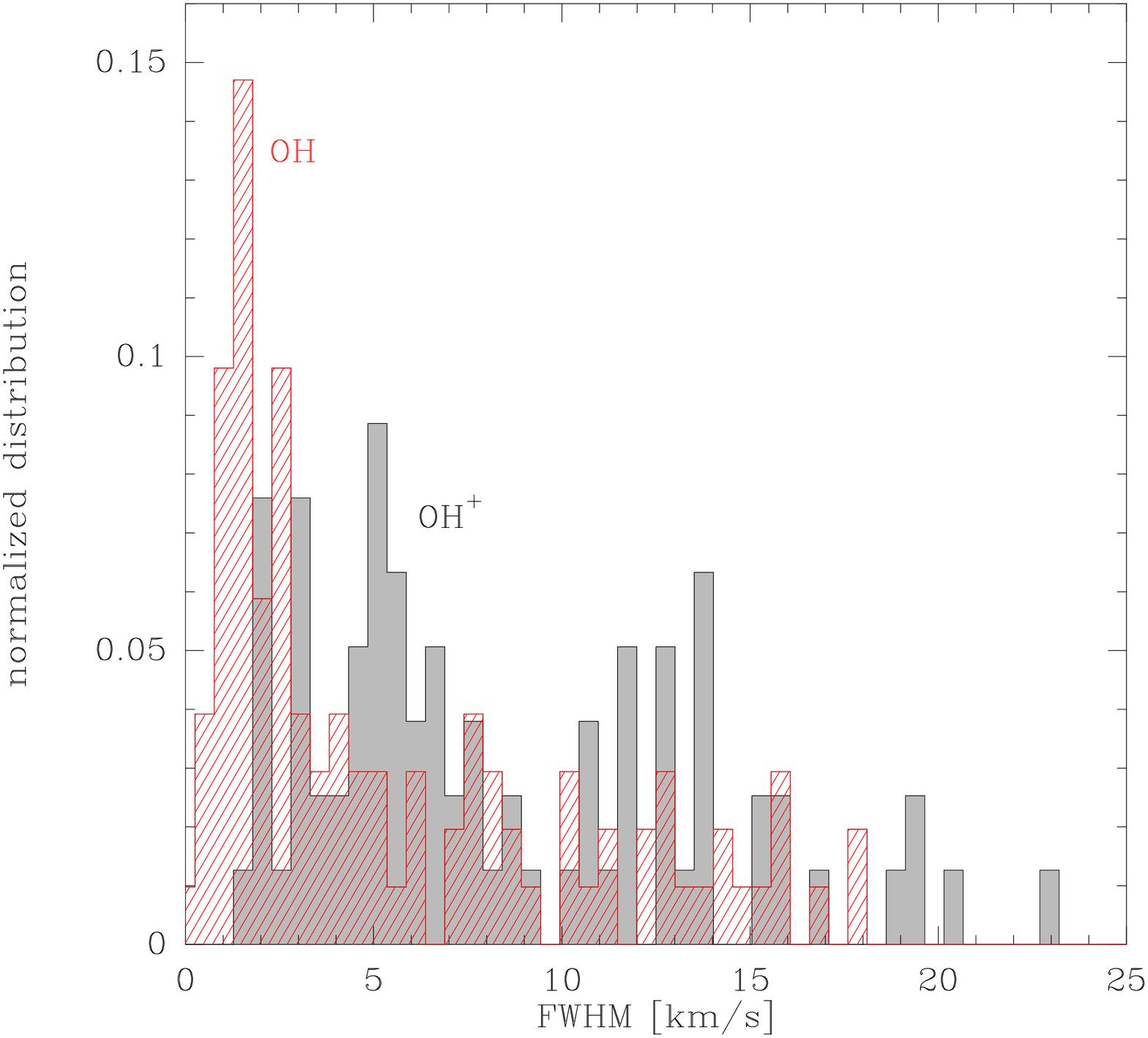}
   \vspace{2mm}
   \caption{Normalized distribution of the velocity widths (FWHM) of Gaussian
            velocity components (red: OH, gray: $\OHP$).}
   \label{fig:distribution_fwhm}
   \end{figure}
   \begin{figure}[hb!]
   \centering 
   \includegraphics[width=\columnwidth]{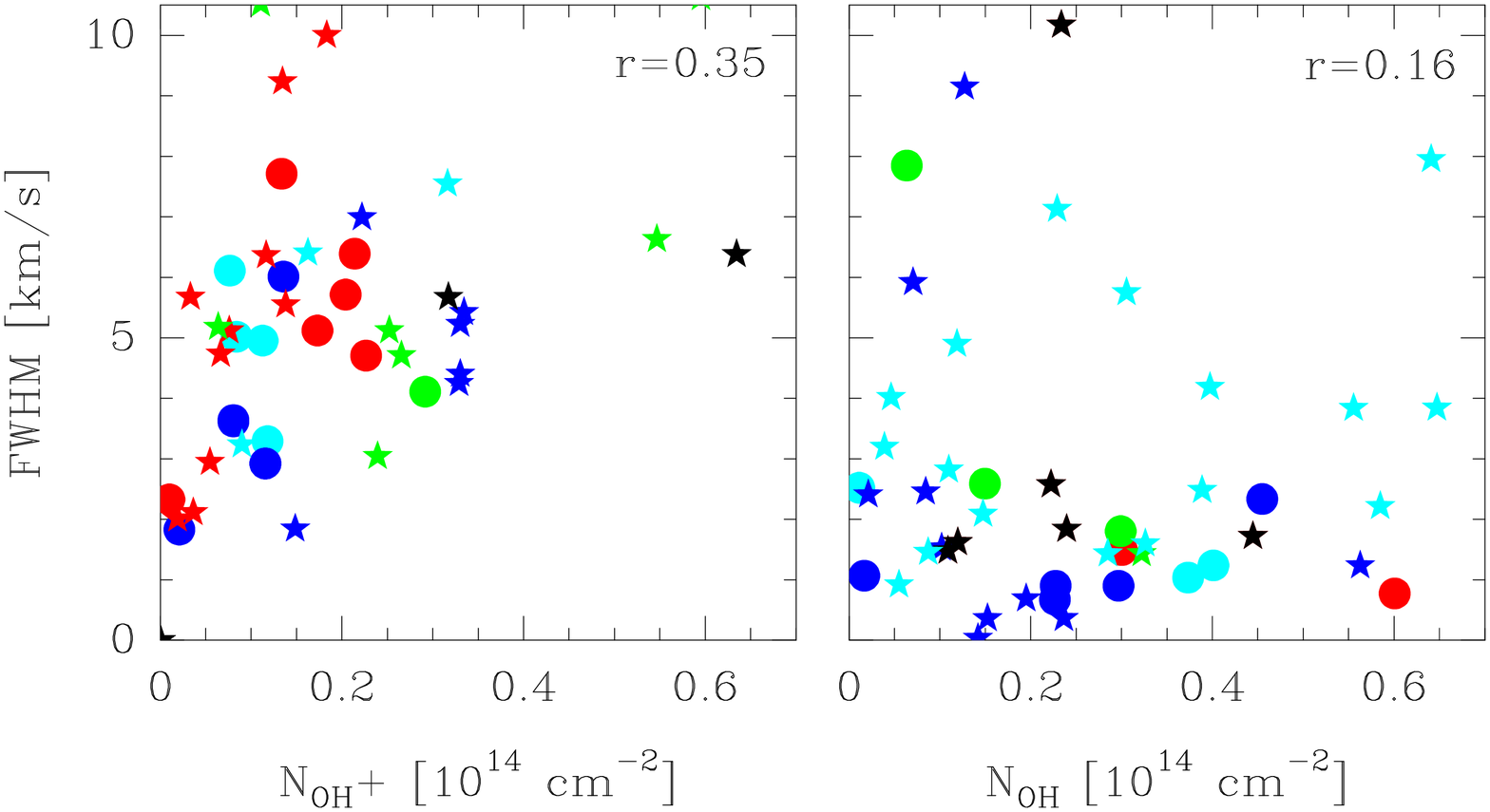}
   \caption{Comparison of line widths (ordinate) with column densities (abscissa)
            of the fitted Gaussian velocity components.
            The circles are for sightlines in the first quadrant, the
            stars for sightlines in the fourth quadrant. Each color
            represents a sightline toward a different target. Left:
            $\OHP$, right: OH.} 
   \label{fig:fwhm_vs_colden}
   \end{figure}
\subsection{Chemistry of OH and OH$^+$}
The median value of the column density of OH in the first quadrant is $3.9\times 10^{14}$~cm$^{-3}$ and for $\OHP$ it is $0.68\times 10^{14}$~cm$^{-3}$.
In the fourth quadrant, the median column densities amount to $1.7\times 10^{14}$~cm$^{-3}$ and $0.55\times 10^{14}$~cm$^{-3}$, respectively. Column
densities in excess of $10^{15}$~cm$^{-2}$ (for OH) and above $\sim$$ 10^{14}$~cm$^{-2}$
(for $\OHP$) are rather the exception. The median $N(\OH)/N(\OHP)$ ratio over all sightlines and velocity components is 3.3.
While the formation of $\OHP$ results from cosmic-ray induced 
reactions involving atomic and molecular hydrogen and atomic oxygen, the bottleneck expected for the formation of
OH via ion-neutral chemistry is the availability of $\HH$. The reaction of $\OHP$ with $\HH$ yields $\HHHOP$, via the two
hydrogen abstraction reactions $\OHP(\HH,\H)\HHOP(\HH,\H)\HHHOP$ (see Appendix \ref{app:C}). Then the dissociative recombination
of $\HHHOP$ yields OH and $\HHO$, with a branching ratio of $\sim$74\% to 83\% in favor of OH \citep[determined by ion storage ring
experiments,][]{2000ApJ...543..764J,2000JChPh.113.1762N}, while less than $\sim$1\% forms \OI. In the following, we attempt to confirm
these predictions, that is, the bottleneck reaction $\OHP$($\HH$,H)$\HHOP$, and the $N(\OH)/N(\HHO)$ ratio. As for the former, a strong anticorrespondence between the column densities of OH and $\OHP$might naively be
expected, where the availability of $\HH$ tips the scales in favor of
OH, while OH$^+$ traces predominantly atomic gas \citep[][further references therein]{2012ApJ...754..105H}. But even if a clear anticorrelation
between OH and $\OHP$ existed, it would be impossible to observe it. On a given sightline several clouds with high and low molecular hydrogen
fractions $f^{\rm N}_\HH = N(\HH)/(2N(\HH)+N($\HI$))$ line up. Even across a single diffuse cloud, the $N(\OH)/N(\OHP)$ ratio is expected to vary
substantially, depending on the degree of self-shielding of $\HH$ against the interstellar UV radiation field. To quantify the anticorrelation, 
we normalized the velocity-specific OH column density with the total OH and $\OHP$ reservoir and obtained an abundance ratio
$r = N_{\rm v}(\OH)/(N_{\rm v}(\OH)+N_{\rm v}(\OHP)$ varying from zero (only $\OHP$, no OH) to one, where all the $\OHP$ abundance is exhausted owing to
the formation of OH and (see below) water. (The normalization chosen here avoids the divergence of the
distribution if OH has no spectral counterpart in $\OHP$.) The resulting distribution (Fig.~\ref{fig:contrast})
indeed shows that these extremes are present in the data, although the second case is by an order of magnitude more
frequent. We suggest
two explanations for this. One reason is that if $f^{\rm N}_\HH$ is too small, $\OHP$ can be efficiently destroyed
by the dissociative recombination with free electrons (Appendix \ref{app:C}), while the formation of $\OHP$ by the
reaction chain $\HP(\O,\H)\OP(\HH,\H)\OHP$ and the secondary, less important path $\HHP(\HH,\H)\HHHP(\O,\HH)\OHP$ become
less efficient (see Appendix \ref{app:D}) because less $\HH$ is available. Another reason is that the fraction $f^{\rm N}_\HH$ is larger
in denser gas (cf. Table~\ref{table:2}) where column densities are higher and absorption features easier to observe.

The anticorrelation between $N({\rm OH})$ and $N(\OHP)$ is expected to increase with the fractional abundance of molecular hydrogen,
$f^{\rm N}_\HH$. Again using HF as surrogate for $\HH$ with $X({\rm HF}) = 1.4\times 10^{-8}$, for W31C and W49N we indeed find
a correlation between the $N({\rm OH})/N(\OHP)$ ratio (Fig.~\ref{fig:contrast} shows that divergence of this ratio is excluded),
with coefficients $\rho = 0.42$, 0.02 and 0.43 toward W31C, G34.26, and W49N, respectively, and false-alarm probabilities of 6\%, 94\%, and
3\% (Fig.~\ref{fig:bottleneck}, again, only data points with relative error <20\% are retained). We note that qualitatively similar
but less significant correlations can be deduced using \OI instead of \HI and $\HH$, in agreement with the results shown in Sect. 4.1.
The lack of a significant correlation toward G34.26 is most likely explained by the few data points available on this
relatively short line of sight (1.56~kpc, see Appendix \ref{app:A}). As for the sightline to W31C, the significant correlation has to be
interpreted with care: The background source is
located in the 3~kpc arm where the density of Galactic free electrons is an order of magnitude above its value in the solar neighborhood
\citep{2001AJ....122..908G}. In such an environment the $n(\HH)/n(e^-)$ ratio falls short of $\sim$100 (cf. Table~\ref{tab:reactionRates}),
so that neither $\HHOP$ nor the subsequent products can be formed owing to the dissociative recombination of $\OHP$. Furthermore, the
complex gas kinematics in the 3~kpc arm \citep[further references therein]{2014ApJ...781..108S} and the resulting confusion due to the
sightline crowding mentioned above add to the complexity, and drawing more quantitative conclusions proves to be difficult. However, given
that the correlation toward W49N is also significant, it seems fair to say that our results confirm the importance of the reaction
$\OHP$($\HH$,H)$\HHOP$ and conclusions that $\OHP$ is rather associated with diffuse, atomic gas \citep{2010A&A...518L.110G,2010A&A...521L..10N},
while the OH seen in unsaturated absorption is located in diffuse molecular and translucent clouds.
The underabundance of $\OHP$ in the latter with respect to OH was also observed in UV spectroscopy \citep{2010ApJ...719L..20K}.

We derived OH abundances $X(\OH) \sim$$10^{-8}$ up to $\sim$$10^{-6}$, and $\OHP$ abundances from $\sim$$10^{-8}$ to $\sim$$10^{-7}$.
The smallest and largest OH abundances were encountered on the sightline to G10.47, in the 135~\kms arm and the Sagittarius arm, respectively.
Both values are somewhat uncertain because only CH is available as $\HH$ tracer and, for the 135~\kms arm, the same caveat holds
as for the
environment of W31C.
The OH abundances for the remaining sightlines agree reasonably well (i.e., within error bars) with the values predicted by 
\citet{2014ApJ...787...44A}, who modeled the chemistry in diffuse clouds, including the time-dependence
of the ortho-to-para ratio of $\HH$, $\HHP$ and $\HHHP$, gas-grain interactions and grain surface reactions.
They obtained $X(\OH) = (0.3-1.6)\times 10^{-7}$, based on the chemical model underlying the Meudon PDR
code \citep{2006ApJS..164..506L}, which includes reactions on the surface of dust grains.
The models of \citet{2014ApJ...787...44A} do not include the endothermic reactions of warm chemistry,
triggered by turbulent dissipation regions (TDRs) or slow shocks, while the precision of our OH abundance determinations is not good enough
to exclude them. An additional piece of information is the production of water by the following reaction chain: $\O(\HH,\H)\OH(\HH,\H)\HHO$
with energy barriers of 2980~K and 1490~K. This warm chemistry was investigated by \citet{2012A&A...540A..87G}. According to the authors, the models
with lower rates of turbulent strain and thus dominated by ion-neutral drift yield the best description of the observed abundance
pattern. Thanks to the aforementioned highly endothermic path to the production of OH and $\HHO$, the abundance ratio of $X(\HHO)/X(\OH)$ is
sensitive to the dissipation of turbulence. \citet{2012A&A...540A..87G} obtained a typical ratio of 0.16 for a UV shielding
of $A_{\rm V} = 0.4$~mag and $n_{\rm H} = 100$~cm$^{-3}$. Comparing our OH column densities with those determined for para-water by
\citet{2010A&A...521L..12S}, we obtain for their velocity intervals the correlation shown in Fig.~\ref{fig:water}. The ortho/para ratio
in translucent clouds was determined by \citet{2013ApJ...762...11F} to be 3:1, its equilibrium value in the high-temperature limit. The
resulting abundance ratio is $X(\HHO)/X(\OH) = 0.28$, which is
reasonably similar to the value for chemistry driven by turbulence.
\begin{table*}[ht!]
\caption{Abundances of OH and of $\OHP$ with respect to \OI. Velocity intervals and spiral arm assignments as in Table~\ref{tab:results}$^{(a)}$.}
\label{tab:oxygen}
\begin{tabular}{lcccccc} 
\hline\hline \\
 & $T_{\rm c}^{(b)}$& $\upsilon_{\rm min}, \upsilon_{\rm max}$ & $N$(\OI)              & $f^{\rm N}_\HH\,^{(c)}$ & $\underline{N(\OH)\,\,}\,^{(c)}$ \\
 & [K]                & [\kms]                                 & [$10^{18}$~cm$^{-2}$] &                       & $N(\OHP)\,\,\,\,\,$            \\
\hline
 \vspace{0.5ex}\\
G10.62 & 2.6 & (7,15)  & $1.71 \pm 0.02$ & $ 0.210\pm 0.004 $ & $1.84  \pm 0.07$  \\
       &     & (42,55) & $2.02 \pm 0.04$ & $ 0.144\pm 0.005 $ & $5.54  \pm 0.43$  \\
G34.26 & 3.0 & ( 0,15) & $1.62 \pm 0.02$ & $ 0.133\pm 0.004 $ & $3.87  \pm 0.49$  \\
       &     & (15,30) & $1.78 \pm 0.02$ & $ 0.079\pm 0.002 $ & $<0.004$ \\
       &     & (38,46) & $1.19 \pm 0.01$ & $ 0.090\pm 0.002 $ & $1.69  \pm 0.03$  \\
W49N   & 3.5 & (20,30) & $1.33 \pm 0.02$ & $ 0.145\pm 0.003 $ & $0.82  \pm 0.06$  \\
       &     & (30,45) & $5.45 \pm 0.11$ & $ 0.271\pm 0.004 $ & $1.68  \pm 0.02$  \\
       &     & (45,60) & $8.0  \pm 2.7 $ & $ 0.227\pm 0.003 $ & $2.34  \pm 0.04$  \\
\hline

\end{tabular}
\tablefoot{$^{(a)}$ With the following exceptions: G10.62 (alias W31C) -- The $\upsilon_{\rm lsr} = (65,80)$~\kms
interval from Table~\ref{tab:results} is not reported because only $\OHP$ was found there. G34.26 -- The
$\upsilon_{\rm lsr} = (62,85)$~\kms interval from Table~\ref{tab:results} is not reported because the \OI line turns into
emission there. W49N -- The velocity interval for the far-side crossing of the Sagittarius spiral arm is only taken up to
$\upsilon_{\rm lsr} = 60$~\kms because the \OI absorption beyond is saturated. $^{(b)}$ Main-beam temperature of 4.744~THz
continuum (Rayleigh-Jeans scale). $^{(c)}$ Error estimates do not account for systematic uncertainties and are derived
from $\chi^2$ of fits by formal error propagation.}
\end{table*}
   \begin{figure}[ht!]
   \centering 
   \includegraphics[width=\columnwidth]{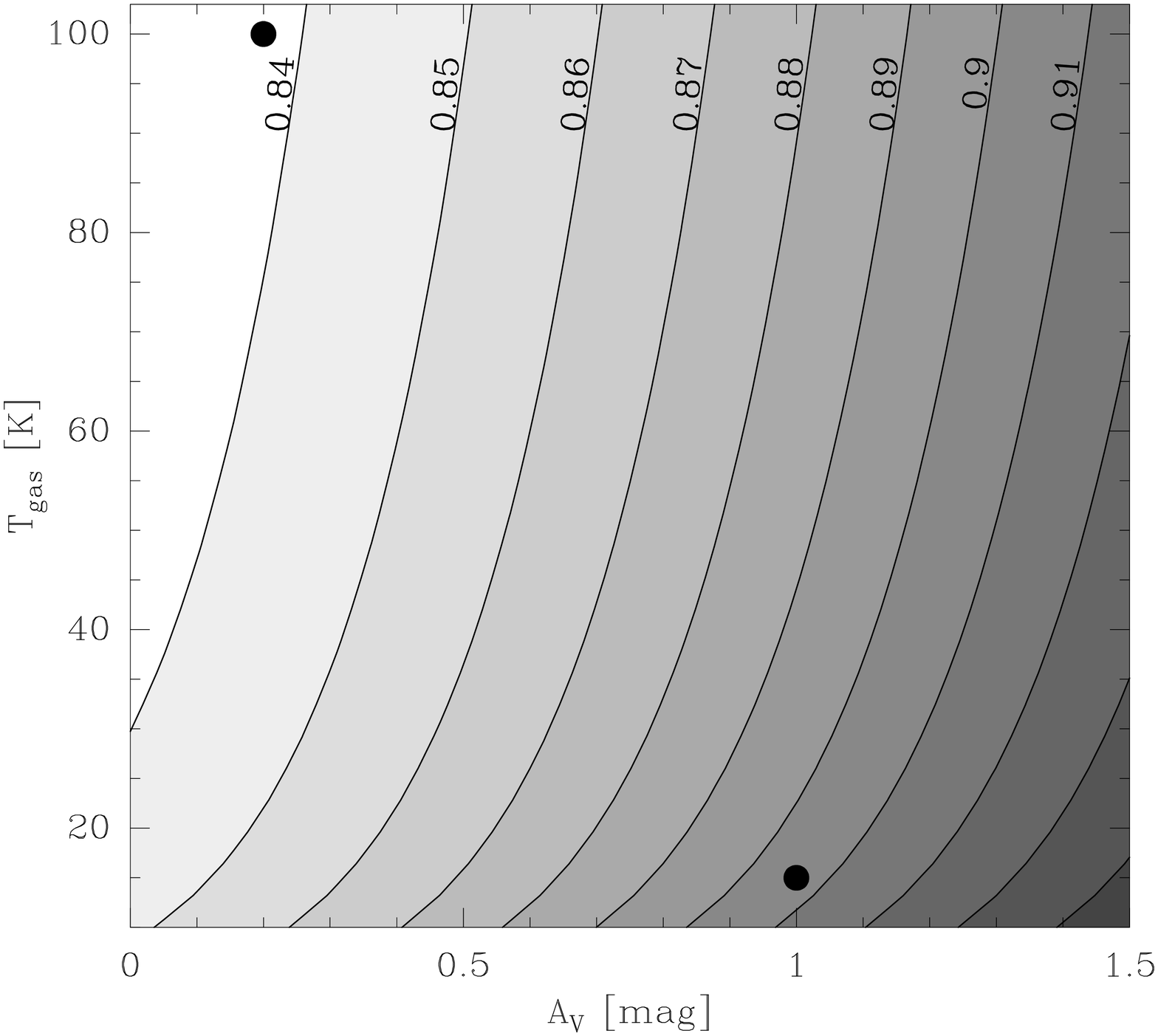}\vspace{2mm}
   \caption{Branching ratio (contour label) for the dissociative recombination of $\HHHOP$ to OH and
$\HHO$ as a function of visual extinction, $A_{\rm v}$ , and gas temperature for an
ortho-to-para ratio of $\HHO$ and of $\HHOP$ of 3:1. Column densities of $\HHO$ are from 
\citet{2010A&A...521L..12S,2015ApJ...806...49S}. \citep{2000JChPh.113.1762N}. The two black dots are at the parameters for models 1 and 2.
For details see text.}
   \label{fig:branching_ratio}
   \end{figure}
\subsubsection{Branching ratio of the dissociative recombination of $\HHHOP$}
We finally attempt to determine the branching ratio of the dissociative recombination of $\HHHOP$, forming OH and $\HHO$. If
the chemistry is dominated by cold ion-neutral reactions, the ratio between $X(\HHO)/X(\HHHOP)$ and $X(\OH)/X(\HHHOP)$ should
reflect the measured branching ratios of 74 to 83\% \citep{2000ApJ...543..764J,2000JChPh.113.1762N}. We adopt the following
phenomenological description of the underlying chemistry, with the rate equations
\begin{eqnarray}
\frac{dn_\OH}{dt}  & = & \Gamma_\OH+\beta\gamma_{\rm DR}n_{\HHHOP}+\gamma_{\rm PD}n_{\HHO}-\Lambda_\OH n_{\OH}\,, \nonumber \\ 
\frac{dn_\HHO}{dt} & = & \Gamma_{\HHO}+(1-\beta)\gamma_{\rm DR}n_{\HHHOP}-(\Lambda_{\HHO}+\gamma_{\rm PD})n_{\HHO}\,,
\label{eq:rateEquations}
\end{eqnarray}
where $\gamma_{\rm DR}$ and $\gamma_{\rm PD}$ are the rates for the dissociative recombination of $\HHHOP$ and, respectively,
the photodissociation of $\HHO$. Gains and losses in the populations of OH and $\HHO$ that are not due to these two processes
are denoted $\Gamma$ and $\Lambda$, respectively. The branching ratio for the dissociative recombination of $\HHHOP$ is denoted
$\beta$. $\HHHOP$ is difficult to observe, and only one measurement is available
\citep[in the $(55,75)$~\kms interval toward W51e2][]{2015ApJ...800...40I}. We therefore used $\HHOP$ instead, with the working
hypothesis that its abundance is proportional to that of $\HHHOP$. For equilibrium chemistry, this assumption is possibly justified.
From our data and those published by \citet{2010A&A...521L..12S,2015ApJ...806...49S} for $\HHO$ and by \citet{2015ApJ...800...40I} for $\HHOP$
we derive significant linear correlations between the column densities of OH and $\HHOP$, and between those
of $\HHO$ and $\HHOP$, with correlation coefficients of, respectively, 0.98 (false-alarm probability $< 1$\%) and 0.87 (false-alarm
probability 5\%). Using Eq. \ref{eq:rateEquations}, the ratio of the slopes $b_{\OH}$ and $b_{\HHO}$ derived from the linear
regression analysis is 
\begin{equation}
\frac{b_\OH}{b_\HHO} = \frac{\gamma_{\rm PD}+\beta\Lambda_\HHO}{(1-\beta)\Lambda_\OH}
\label{eq:ratioOfSlopes}
.\end{equation}
Identifying photoionization and photodissociation as main loss channels for the abundances of OH and $\HHO$
(cf. Tab.~\ref{tab:reactionRates}), we find
\begin{align}
\gamma_{\rm PD}[{\rm s}^{-1}]&= 7.5\cdot 10^{-10} \exp(-1.70A_{\rm v})\,, \nonumber\\
\Lambda_\HHO[{\rm s}^{-1}]&= 3.1\cdot 10^{-11} \exp(-3.90A_{\rm v})+4.8\cdot 10^{-11} \exp(-2.20A_{\rm v}) \,,\nonumber\\
\Lambda_\OH[{\rm s}^{-1}]&= 2.2\cdot 10^{-11} \exp(-4.05A_{\rm v})+3.9\cdot 10^{-10} \exp(-1.70A_{\rm v}) \nonumber \\
&\,\,\,\,\,+2.9\times 10^{-9}(T/300 {\rm K})^{-0.33} n_{\rm C^+}\,[{\rm cm^3}]\,. 
\end{align}
The first term on the right-hand side of the loss rate for $\HHO$ is due to its photoionization, the second one
due to the photodissociation of $\HHO$ to \OI. Likewise, the first two terms on the right-hand side of the loss rate of OH
are due to photoionization and to photodissociation, respectively, while
the last contribution to the loss channels of OH comes from the ion-neutral reaction ${\rm C}^+(\OH,\H){\rm CO}^+$, for which we assumed
a C$^+$ density of $n_{\rm C^+} \sim$0.01~cm$^{-3}$, in agreement with the models shown in Appendix \ref{app:D} and Table~\ref{table:2}. The
deduced branching ratios are shown in Fig.~\ref{fig:branching_ratio} for visual extinctions and gas temperatures in the range of our models.
The underlying column densities are shown in Table \ref{tab:colden_water_etc}.
Up to $A_{\rm V} = 1.0$~mag the branching ratio depends only weakly on temperature. For the regime of our models we obtain
a branching ratio of 0.84 - 0.91. This is compatible with the value of 0.83 determined by \citet{2000JChPh.113.1762N}. We note that the
p-$\HHO$ column densities of \citet{2010A&A...521L..12S} were again corrected by the ortho/para ratio of 3:1, and the
same ortho/para ratio for $\HHO$ and $\HHOP$ was assumed.
\begin{table}[h!]
\caption{Column densities of p-$\HHO$, $\HHOP$ and OH used in the determination of the branching ratio.}
\label{tab:colden_water_etc}
\begin{tabular}{cccc}
\hline\hline
$\upsilon_{\rm min}, \upsilon_{\rm max}$ & $N(\OH)$ & $N(p-\HHO)^{(\rm a)}$ & $N(p-\HHOP)^{(\rm b)}$ \\
$[$km\,s$^{-1}]$ & \multicolumn{3}{c}{[$10^{13}$~cm$^{-2}$]}    \\
\hline
\multicolumn{4}{c}{W49N} \\
\hline
$30-50$ & $4.0\pm 0.1$   & $2.2\pm 0.8$    & $5.3\pm 0.4$ \\
$50-78$ & $3.9\pm 0.1$   & $3.4\pm 0.9$    & $2.4\pm 0.2$ \\
$67-71$ & $0.17\pm 0.04$ & $0.15\pm 0.03$  & $0.8\pm 0.1$ \\
\hline
\multicolumn{4}{c}{W51} \\
\hline
$0-10$  & $11.4\pm 0.9$ & $0.62\pm 0.05$ & $0.6\pm  0.1$  \\
$10-20$ & $2\pm 1$      & $0.04\pm 0.01$ & $0.26\pm 0.03$ \\
$42-47$ & $6.0\pm 0.3$  & $0.54\pm 0.07$ & $0.99\pm 0.05$ \\
\hline
\end{tabular}
\vspace{0.5ex}\\
\tablefoot{$^{(\rm a)}$ \citet{2010A&A...521L..12S};$^{(\rm b)}$ \citet{2015ApJ...800...40I}.}
\end{table}
\subsection{Gas dynamics and Galactic structure}
In the remainder of this section we briefly address aspects of Galactic structure and gas dynamics. If distance ambiguities 
at a given velocity can be ruled out for example by measurements of maser parallaxes in the continuum sources \citep{2014ApJ...783..130R}, it
is possible to localize the absorbing spiral arm. The absorption may be caused by a single cloud, but more likely by
a blend of several clouds. We note that the absorption features seen in OH tend to be narrower than those observed in $\OHP$, as shown in the
distribution of the respective line widths (Fig.~\ref{fig:distribution_fwhm}). The narrowest line widths may thus be due to absorption in a single
cloud. In the more likely case of a blend of several clouds the largest line widths are due to large velocity gradients
occurring in certain spiral arm crossings. Mild shocks dissipating interstellar turbulence,
with slow to moderate velocities \citep{2013A&A...550A.106L}, may also produce large line widths. The distribution of absorption line widths
indeed suggests a tail above $\sim$10~\kms, with the bulk of fitted profiles at lower widths. 
MHD simulations show that the velocity dispersion along the line of sight increases with column density,
both in super- and sub-Alfvénic models, but the correlation between these two quantities is stronger in the supersonic than in the
subsonic regime (\citealp{2009ApJ...693..250B}, see also \citealp{1998ApJ...504..300P}). For $\OHP$ the correlation coefficient is 0.35 and
the correlation is significant at the 5\% level, that is, 12\% of the variance observed in the line widths is probably due to
the expected correlation between column density and line width. For OH, the correlation is much weaker due to the larger number of narrow
velocity components, regardless of column density. With a coefficient of 0.16, no correlation between column density and line width is found
with a false-alarm probability of $\le 5$\%. Dissipation of turbulence would tend to flatten the correlation. At present we cannot confirm
whether the narrow OH absorption features, originating from diffuse molecular and translucent clouds, are a direct consequence of dissipation
of turbulence. However, it seems fair to say that the observed distribution of line widths is a genuine manifestation of the interplay between
interstellar turbulence and the association of $\OHP$ with more diffuse, atomic and of OH with denser, molecular gas.
The higher contrast of arm to interarm seen in OH absorption compared to that observed in OH$^+$ is yet another manifestation of the
anticorrespondence between the two spieces: In Figs.~\ref{fig:colden1stqu} and \ref{fig:colden4thqu}, the groups of enhanced column
densitities corresponding to different spiral arm crossings are separated more clearly in OH than in OH$^+$. A striking example
is the sightline to W49N. In the $(20,30)$~\kms interval that we assigned to the interarm region between
the Perseus and Sagittarius arms, the ratio between the respective OH column densities is 5 for OH, but only 2 for $\OHP$. Likewise,
\citet{1998ApJ...502..265H} found for the contrast between the outer Perseus arm and the interarm gas a ratio of $2.5$ in atomic gas
traced by \HI, but a ratio of $28$ in molecular gas, traced by CO.
We finally note that the characteristic time for crossing a spiral arm, and
therefore for the compression of gas by the corresponding density wave, is $\sim$80~Myr in the solar neighborhood, 
15~Myr at $R_{\rm G} = 4.5$~kpc, and 68~Myr at $R_{\rm G} = 7.2$~kpc, assuming that the diffuse gas rotates with the
same drift speed as the stellar clusters for which these crossing times were derived \citep{2007MNRAS.376..809G}.
The timescale on which chemical models of diffuse and translucent clouds reach equilibrium abundances is $\sim$10~Myr
\citep[e.g.,][]{1992MNRAS.258..377H}. It can therefore not be ruled out that the chemistry in the observed clouds, at least for the smaller
galactocentric radii, is not in equilibrium. The order-of-magnitude variation found by us in the fractional
abundances of OH and $\OHP$ would agree with this statement, as would the lack of a significant correlation between the X(OH)/X($\OHP$)
ratio and the molecular hydrogen fraction toward W31C, at low $R_{\rm G}$ (Fig.~\ref{fig:bottleneck}).
   \begin{figure}[h!]
   \centering 
   \includegraphics[width=\columnwidth]{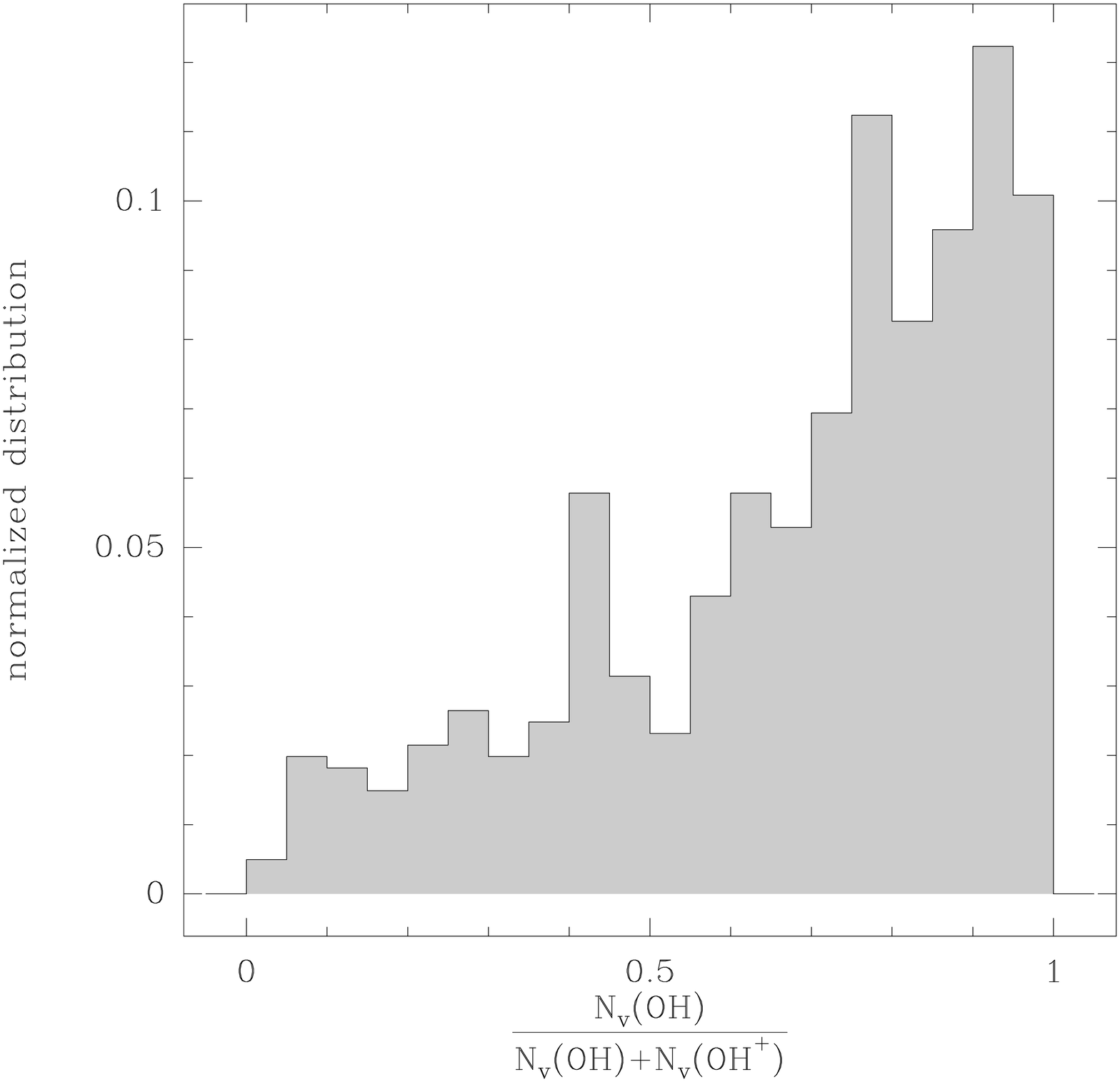}
   \vspace{2mm}
   \caption{Normalized distribution of the velocity-specific
            column density of OH, $N_{\rm v}(\OH)$, with respect to $N_{\rm v}(\OH)+N_{\rm v}(\OHP)$.}
   \label{fig:contrast}
   \end{figure}
   \begin{figure}
   \centering 
   \includegraphics[width=\columnwidth]{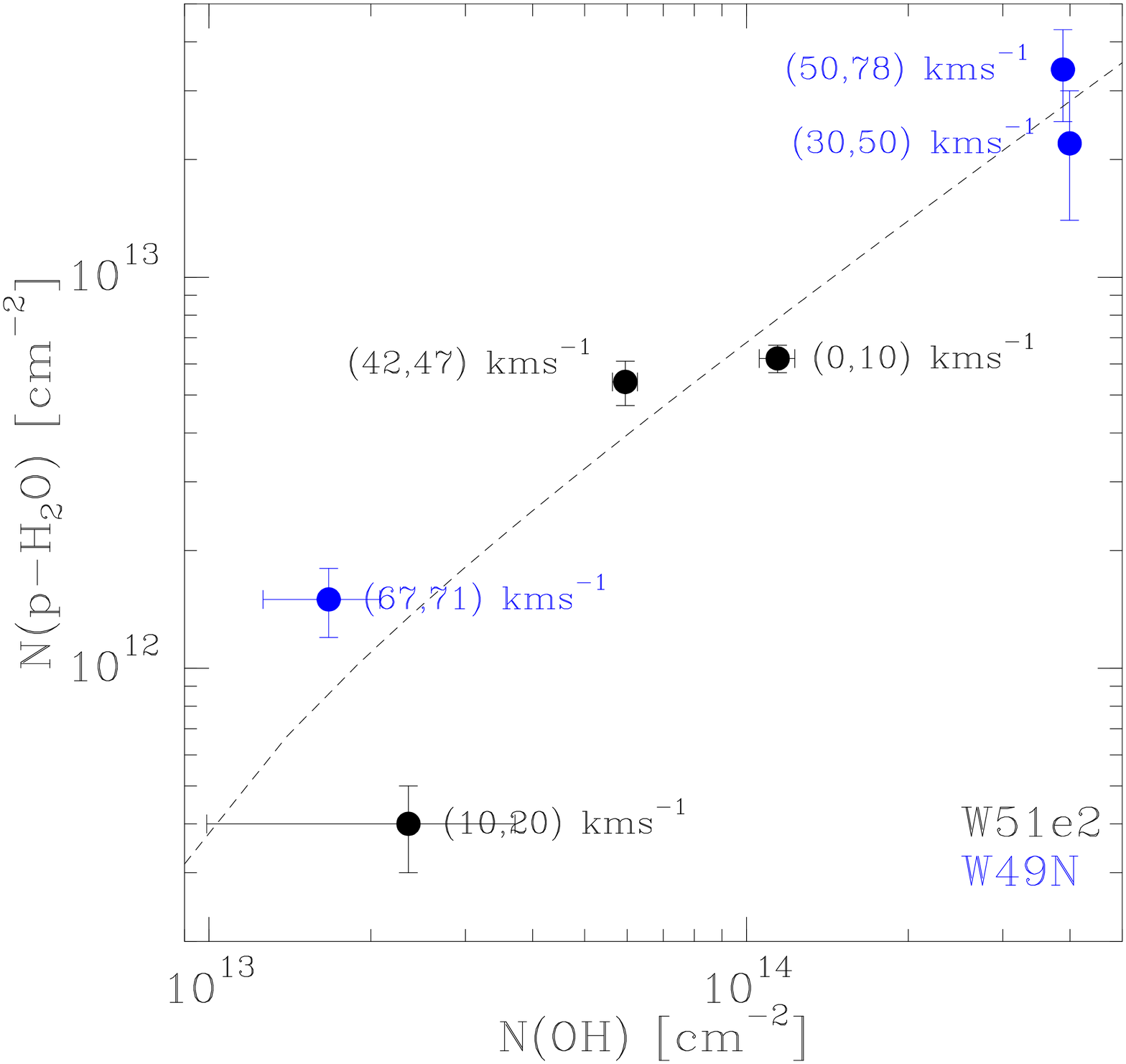}\vspace{2mm}
   \caption{Correlation between the column densities of OH (this work) and $p-\HHO$ \citep{2010A&A...521L..12S},
            for clouds on the sightlines to W51e2 (black) and W49N (blue). The dashed line shows the linear
            regression between these quantities. The velocity intervals used to derive the column densities are indicated.
            $N(\HHO)$ values are from \citet{2010A&A...521L..12S} and \citet{2015ApJ...806...49S}.}
   \label{fig:water}
   \end{figure}
\section{Conclusions}
\label{sec:conclusions}
We presented the first dedicated survey of absorption spectra of OH, $\OHP$ and \OI in diffuse interstellar clouds with
adequate velocity resolution, allowing us to separate the observed features into several spiral arm crossings.
We conclude with a summary of our most important findings.\begin{enumerate}
\item The \OI absorption was observed for three sightlines with unprecedented spectral resolution. We showed that
a significant correlation exists between the column density of \OI and that of the total (atomic and molecular) hydrogen
reservoir. This is because \OI is only slowly removed from the gas phase as diffuse atomic clouds evolve toward diffuse
molecular and translucent clouds. We found that the sightline-averaged \OI abundances toward W31C, G34.26, and W49N are confined 
to a narrow range of 3.1 to $3.5\times 10^{-4}$, which agrees reasonably well with earlier measurements \citep[e.g.,][]{2001ApJ...561..823L}
and below the abundance of $(5.75 \pm 0.4)\times 10^{-4}$ in the B-type stars of nearby OB-associations \citep{2008ApJ...688L.103P}. If
we take the latter abundance as reference standard, the difference with our abundance measurement can be attributed to the depletion
of \OI into oxygen-bearing molecules, ices, and dust. However, our result does not exclude the presence of so far
unidentified carriers of \OI \citep{2009ApJ...700.1299J,2010ApJ...710.1009W}. We finally note that with the distance measurements of
\citet{2014ApJ...783..130R} and toward suitably chosen sightlines, it should be possible to determine a galactocentric \OI abundance
gradient from first principles, without the non-LTE modeling required for excited oxygen lines, and free of the opacity limitations
of UV spectroscopy.
\item The column densities of OH are loosely correlated with those of $\OHP$, reflecting the spiral arm crossings on
the observed sightlines. The arm-to-interarm contrast is stronger in OH than in OH$^+$. The two radicals can coexist, but
there are more cases where OH exists without $\OHP$ than the other way around. We attribute this finding to the dissociative
recombination of $\OHP$ occurring if there is not enough $\HH$ to form OH and $\HHO$.
\item The $\OH/\OHP$ ratio increases as a function of the molecular fraction $f^{\rm N}_\HH = N(\HH)/(N(\H)+2N(\HH))$.
We interpret this as an indication of the importance of the bottleneck reaction
$\OHP(\HH,\H)\HHOP$ for the formation of OH and $\HHO$ in cold, ion-neutral driven chemistry, while our $N(\HHO)/N(\OH)$ ratios
do not rule out the endothermic reaction pathways of warm chemistry in turbulent dissipation regions and slow shocks.
\item We estimated the branching ratio for the dissociative recombination of $\HHHOP$ into $\HHO$ and OH and found that a
range of 84 to 91\% of the available $\HHHOP$ yields OH. This compares well with laboratory measurements, which yield 74 to 83\%.
\item The line widths of the velocity components of the $\OHP$ absorption are significantly correlated (at the 5\% level)
with the column density, as expected from models of MHD turbulence. For OH, there is no such correlation, owing to the
frequent occurrence of narrow absorption lines, irrespective of column density. It can only be speculated whether the latter
finding is to be interpreted as a sign of dissipation of turbulence.
\end{enumerate}
The uncertainties in our analysis are mainly due to velocity crowding on the sightline and to uncertainties regarding the abundances
of the $\HH$ tracers. Only the ability of GREAT to observe the ground-state absorption of OH and \OI at the
required spectral resolution and at frequencies above those covered by HIFI has allowed us to reduce the sightline confusion to an unavoidable minimum and to
determine column densities from first principles. In summary, it seems fair to say that our findings further constrain the importance of cold ion-neutral
chemistry in the diffuse interstellar medium of the Galaxy without excluding endothermic neutral-neutral and grain surface reactions. 
\begin{acknowledgements}
Based in part on observations made with the NASA/DLR Stratospheric
Observatory for Infrared Astronomy. SOFIA Science Mission Operations are
conducted jointly by the Universities Space Research Association, Inc., under
NASA contract NAS2-97001, and the Deutsches SOFIA Institut under DLR contract
50 OK 0901. We {\sc great}fully acknowledge the support by the observatory staff
and the careful examination of this work by an anonymous referee.
The kinetic data used in our study were downloaded from the online database KIDA
(Wakelam et al. 2012, http://kida.obs.u-bordeaux1.fr).
\end{acknowledgements}
%

%
\begin{appendix}
\section{Comments on individual sightlines}
\label{app:A}
The next two paragraphs introduce the observed sightlines, that
is, their spiral arm crossing, and provide 
a phenomenological description of the absorption spectra.
\subsection{First quadrant}
{\bf G10.47 -} The sightline toward G10.47, located at 8.5~kpc in the connecting arm
\citep{2014ApJ...781..108S}, exhibits a relatively broad absorption spectrum. The features are from the
near side of the Sagittarius arm, the Scutum arm, the near 3~kpc arm, and
the Galactic bar. By consequence, the column density profiles of OH and $\OHP$
extend from -20 to 160~\kms with a rough correspondence between OH and $\OHP$
and generally narrower features in OH. There are three major components, at velocities
redshifted with respect to the hot core. They probably belong to the 135~\kms arm
\citep[][further references therein]{2015MNRAS.446.4186S}.
~\\[1.5ex]
{\bf G10.62 (W31C) -} G10.62 is located at a distance of 4.95~kpc in the 3~kpc arm \citep{2014ApJ...781..108S}. The sightline to it crosses the
near sides of the Sagittarius and Scutum arms, with projected velocities
ranging from 0 to 50~\kms. The column densities of OH and $\OHP$ are roughly
correlated with each other, except for an $\OHP$ feature at 60 to 80~\kms
without a corresponding OH feature. Likewise, another local maximum of the
$\OHP$ column density at about 10~\kms is not associated with a corresponding
maximum in the distribution of OH features and represents another remarkable
example of an anticorrespondence of the two species.
~\\[1.5ex]
{\bf G34.26 -} G34.26 is located in the Sagittarius spiral arm, which is the only one crossed 
by this sightline. Its distance of 1.56~kpc is based on the assumption that
G34.34 \citep[$\HHO$ maser parallax,][]{2011PASJ...63..513K} belongs to the same complex.
OH is seen from 45 to 60~\kms and, after a local mininum, from 60 to 75~\kms. $\OHP$ is only observed in the
first interval. Another group of OH at 0 to 15~\kms is framed by local maxima of $\OHP$, again in
anticorrelation. Interestingly, $\OHP$ is seen along the whole sightline, that is, in the interval
from -5 to +75~\kms.
~\\[1.5ex]
{\bf W49N -} W49N \citep[11.1~kpc distance,][]{2013ApJ...775...79Z} shows absorption in three groups, whose separation is clearer in
OH than in $\OHP$. The three groups with velocities from 30 to 70~\kms arise from a sightline grazing the
Sagittarius arm, while the gas from $-10$ to 20~\kms is located in the 
Perseus arm in which W49 is located. Between these two velocity intervals, $\OHP$ is seen with much
less contrast (with respect to the column densities in the two velocity intervals) than OH, which
leads us to assume that the interval from 20 to 30~\kms contains interarm gas located close to the
Perseus arm.
~\\[1.5ex]
{\bf W51e2 -} The sightline to W51e2, at 5.4~kpc distance \citep{2010ApJ...720.1055S} closer than W49N
but at similar Galactic longitude, tangentially follows the Sagittarius arm, with velocities from 40 to 75~\kms
and in the 0 to 10~\kms interval. Between these velocity intervals we detected no OH but $\OHP$. 
\subsection{Fourth quadrant}
~\\[1.5ex]
{\bf G327.29 -} For this hot core \citet{2011MNRAS.417.2500G} provided no distance estimate. The
in projection nearby G329.031 (like G327.29 below the Galactic plane) was shown
to be at 3.2~kpc distance. This would place the target in the Crux spiral arm,
in agreement with the velocities of its associated OH and CH$_3$OH masers. \citet{2015arXiv150300007W} determined
the distance to 3.1~kpc, using the rotation curve of \citet{1993A&A...275...67B} and solving the 
distance ambiguity by means of \HI self-absorption.
The OH column in the $(-10 \pm 10)$~\kms interval coincides with the crossing
of the Carina arm, the first one on the sightline from the observer to G327.29.
The velocity distributions of OH and $\OHP$ are correlated, but with markedly 
different widths.
~\\[1.5ex]
{\bf G330.95 -} Its distance of 4.7~kpc \citep{2011MNRAS.417.2500G} to 5.7~kpc \citep{2015arXiv150300007W}
locates this target in the Norma spiral arm. The Carina and Crux arms 
are traced at about 0~\kms and in the $(-50 \pm 10)$~\kms interval, respectively.
The absorption in the interval at $(-90\pm 10)$~\kms is from the Norma arm and the
environment of the hot core.  The much wider distribution of $\OHP$ is only loosely
correlated with that of OH. Notably, the crossing of the Carina arm only shows in $\OHP$ absorption,
but not in OH.
~\\[1.5ex]
{\bf G332.83 -} The distance to G332.83 is undetermined in \citet{2011MNRAS.417.2500G}. The in projection
nearby region G332.81 is located at 11.7~kpc distance in the far side of the Crux spiral arm. If we were to adopt
this distance, the sightline would cross the same spiral arms as G330.95 but twice the Crux arm: The first
crossing is at $\sim$4~kpc, the second one close to the target itself. However, the resulting tangential crossing
of the Norma arm at $\sim$100~\kms, see also \citet{2008AJ....135.1301V} (his Fig.~2) leaves no trace of an
absorption, neither in OH nor in $\OHP$. We use this as an indication that G332.83 is located in the near side
of the Crux arm. Interestingly, \citet{2015arXiv150300007W} also adopted the near-side distance of 3.2~kpc,
based on a spectrophotometric measurement.
~\\[1.5ex]
{\bf G351.58 -} At 5.1~kpc distance \citep{2011MNRAS.417.2500G}, the hot core is located near the beginning of the Norma
or in the near 3~kpc arm; this is also evidenced by its high negative systemic velocity.
\citet{2015arXiv150300007W} corrected this distance to a value of
6.8~kpc (with \HI self-absorption). As for the other sightlines in the fourth quadrant, the gas near zero 
velocities stems from the Carina arm, while the group ranging from $-70$ to
$-40$~\kms belongs to the Crux arm, which is crossed over a length of
$\sim$1.5~kpc. The gap until the $(-100,-90)$~\kms interval is filled by
interarm gas seen in $\OHP$ with a broad velocity distribution, but not
by a measurable column density of OH.
\section{Observational bias in abundance averages }
\label{app:B}
As previously emphasized, absorption spectroscopy against spatially unresolved background continuum sources
samples the column density, and not the mean gas density. Consequently, the determination of relevant abundances,
or abundance ratios, is not straightforward. If such quantities are expected to qualify for conclusions regarding
a given spiral arm crossing, the question of a potential observational bias has to be addressed. Usually, column
density averages across the corresponding velocity interval $(\upsilon_1, \upsilon_2)$ are applied. If $X$ and $Y$ denote
the column densities per unit velocity interval of the two species to be compared, this leads to ratios $r$ like
\begin{equation}
r = \frac{\int_{\upsilon_1}^{\upsilon_2}{X(\upsilon)d\upsilon}}{\int_{\upsilon_1}^{\upsilon_2}{Y(\upsilon)d\upsilon}}
\label{eq:ratio1}
.\end{equation}
The following analysis shows that in practice such ratios are only of limited use if observational bias is to
be avoided. We note that an absorption line profile is mainly decomposed for deconvolution from hyperfine
splitting (and for the sake of internal consistency also for species without hyperfine splitting), but not to identify
individual Gaussian components as single, veritable diffuse clouds. Considering therefore that each velocity within
the interval $(\upsilon_1,\upsilon_2)$ represents a different ensemble of diffuse cloud entities, ratios like
\begin{equation}
r = \int_{\upsilon_1}^{\upsilon_2}{\frac{X(\upsilon)}{Y(\upsilon)}w(\upsilon)d\upsilon}
\label{eq:ratio2}
\end{equation}
are preferred, where $w(\upsilon)$ is a normalized weight function. The most appropriate choice, namely by mass per unit velocity interval, is
also the most inaccessible one, especially in absorption spectroscopy toward unresolved continuum sources in the background.
The ratios defined in Eqs. \ref{eq:ratio1} and \ref{eq:ratio2} are equal only if the weight function is defined
through the column density of the reference species, that is,
\begin{equation}
w(\upsilon) = \frac{Y(\upsilon)}{\int_{\upsilon_1}^{\upsilon_2}{Y(\tilde{\upsilon})d\tilde{\upsilon}}}\,.
\label{eq:ratio3}
\end{equation}
For the most frequently encountered case, that is, for a diffuse cloud whose angular extent exceeds that of the unresolved background
source, such a weighting would depend on the distance $D$ of the absorbing cloud: Assuming that a typical
Galactic hot core extends over a typical size of $L_{\rm hc} \sim 10^{-2}$ to $10^{-1}$~pc and a diffuse cloud over a typical scale
of $L_{\rm dc} \sim 10$~pc, absorption spectroscopy of a diffuse cloud (at distance $D_{\rm dc}$) against a spatially
unresolved continuum source at distance $D_{\rm hc}$ traces a cloud volume fraction of
\begin{equation}
f^{\rm (V)}_{\rm DC} \sim \left ( \frac{L_{\rm hc}}{L_{\rm dc} D_{\rm hc}} \right )^2 D_{\rm dc}^2. 
\end{equation}
If we consider the diffuse cloud as a whole, only a volume fraction of at most $\sim 0.01$\% is traced at
any distance, which is not
necessarily representative for the entire cloud. However, it is conceivable that what is observed as a single diffuse cloud consists
of a blend of several cloudlets of a typical size of $\sim$1\,pc, possibly showing abundance variations among them
\citep[][further references therein]{2007SSRv..130..341L} The measured subvolume now amounts to at most 1\% for a cloudlet close to the
hot core, but decreases to 0.25\% for a cloudlet located half-way to the hot core, leading to a corresponding weighting of column densities although
both cloudlets are equally representative of the abundance study. Such a situation naturally arises due to the velocity amibiguity
encountered in near- and far-side crossings of a given spiral arm. Moreover, the deduced column densities are by no means representative of the total
mass of clouds at a given line-of-sight velocity. We therefore prefer Eq. \ref{eq:ratio2}, with a uniform weight function, $w(\upsilon) \equiv 1$
in the velocity interval $(\upsilon_1,\upsilon_2)$. The rationale behind this apriority is the assumption that each velocity within an interval
corresponding to a given spiral arm crossing is equally representative for the analysis of the chemistry in diffuse clouds.
This weighting requires elimination of velocity components with an insignificant $Y(\upsilon)$, in order to avoid divergence. Here we
used a conservative cutoff of 5$\sigma_{\rm rms}$. This parameter was varied so as to ensure that the derived abundance ratios did not vary by more
than their inherent mainly noise-induced uncertainty. 
\section{Rates of dissociative recombinations, hydrogen abstraction, and photoionization}
\label{app:C}
The production rate for a chemical reaction A+B $\rightarrow$ C+D with activation energy $\epsilon_0$
is given by \citep[e.g.,][]{1995inco.book.....C}
\begin{equation}
n(A)n(B)k(\epsilon) = n(A)n(B) \int_{\epsilon_0}^\infty \sigma(\epsilon)\Phi(\epsilon)d\epsilon
\label{eq:reactionRate}
,\end{equation}
where $\sigma(\epsilon)$ is the cross section and $\Phi(\epsilon)$ the distribution of the relative
energy $\epsilon$ of species A and B. The parametrization of $\sigma(\epsilon)$ (i.e., the steepness of the
energy barrier) leads to the widely used expression
\begin{equation}
k(T) = \alpha \left(\frac{T}{300}\right)^\beta\exp{\left(-\frac{\gamma}{T}\right)}
.\end{equation}
For the reaction chain of hydrogen abstractions $\OHP(\HH,\H)\HHOP(\HH,\H)\HHHOP$ the parameters $\alpha$,
$\beta,$ and $\gamma$ are calculated by means of the {\it Kinetic Database for Astrochemistry} 
\citep{2012EAS....58..287W} for the conditions defined in Table~\ref{table:2}. The resulting
reaction rates are summarized in Table~\ref{tab:reactionRates}.
\begin{table}
\caption{Rate coefficients for the reactions considered in this work (from \citealp{2012EAS....58..287W} for $\HH$ abstraction reactions, and
from \citealp{2012ApJ...754..105H} for photoionization and photodissociation).}
\label{tab:reactionRates}
\begin{tabular}{lcc} 
\hline\hline \\
                    & \multicolumn{2}{c}{Dissociative recombination rates} \\
                    & model 1 (100~K) & model 2 (15~K) \\ 
                    & \multicolumn{2}{c}{$[\mathrm{cm}^3\mathrm{s}^{-1}]$}      \\
\hline
$\OHP(\mathrm{e}^-,\H)\O$                 & $6.58\times 10^{-8}$ & $1.47\times 10^{-8}$ \\
$\HHHOP(\mathrm{e}^-,\H)\OH$   & $5.54\times 10^{-7}$ & $1.43\times 10^{-6}$ \\
$\HHHOP(\mathrm{e}^-,\HH)\HHO$ & $1.91\times 10^{-7}$ & $4.92\times 10^{-7}$ \\
\hline
                    & \multicolumn{2}{c}{$\HH$ abstraction reactions} \\  
                    & \multicolumn{2}{c}{$[\mathrm{cm}^3\mathrm{s}^{-1}]$}      \\
\hline
$\OHP(\HH,\H)\HHOP$   & \multicolumn{2}{c}{$1.1 \times 10^{-9}$}     \\
$\HHOP(\HH,\H)\HHHOP$ & \multicolumn{2}{c}{$6.1 \times 10^{-10}$}    \\
\hline
                      & \multicolumn{2}{c}{Photodissociation rates ($\chi = 1.7$ Habing)} \\
                      & model 1 ($A_{\rm V}=0.2$) & model 2 ($A_{\rm V}=1.0$) \\ 
                      & \multicolumn{2}{c}{$[\mathrm{s}^{-1}]$} \\
\hline
$\HHO(h\nu,\H)\OH$    & $5.3\times 10^{-10}$ & $1.4\times 10^{-10}$ \\ 
$\HHO(h\nu,\HH)\O$    & $3.1\times 10^{-11}$ & $5.3\times 10^{-12}$ \\ 
$\OH(h\nu,\H)\O$       & $2.8\times 10^{-10}$ & $7.1\times 10^{-11}$ \\ 
\hline
                      & \multicolumn{2}{c}{Photoionization rates ($\chi = 1.7$ Habing)} \\
                      & model 1 ($A_{\rm V}=0.2$) & model 2 ($A_{\rm V}=1.0$) \\ 
                      & \multicolumn{2}{c}{$[\mathrm{s}^{-1}]$} \\
\hline
$\HHO(h\nu,\mathrm{e}^-e)\HHOP$ & $1.4\times 10^{-11}$ & $6.3\times 10^{-13}$ \\ 
$\OH(h\nu,\mathrm{e}^-)\OHP$    & $9.8\times 10^{-12}$ & $3.8\times 10^{-13}$ \\ 
\hline 
\end{tabular}
\end{table}
The rate coefficients of the two dissociative recombinations differ by a factor 1.8, independently of temperature (within
the range under consideration). Without a chemical model at hand (see Appendix \ref{app:D}), the actual production rates are difficult
to estimate. The measured abundance ratio $X(\OHP)/X(\HHOP)$ of up to 10 \citep{2015ApJ...800...40I} reflects the result of this
chemistry rather than its initial conditions. The by three orders of magnitude larger rate coefficient for the dissociative recombination
of $\HHHOP$ is due to the large cross section for this reaction and does not imply a correspondingly higher production rate of
$\HHO$ and OH, given the fraction of free electrons (typically $10^{-4}$ in diffuse atomic and molecular gas, and $10^{-5}$ and less in
translucent clouds where it is maintained by cosmic ray ionization).
\section{Simple single-layer chemical model}
\label{app:D}
   \begin{figure}[ht!]
   \centering 
   \includegraphics[width=\columnwidth]{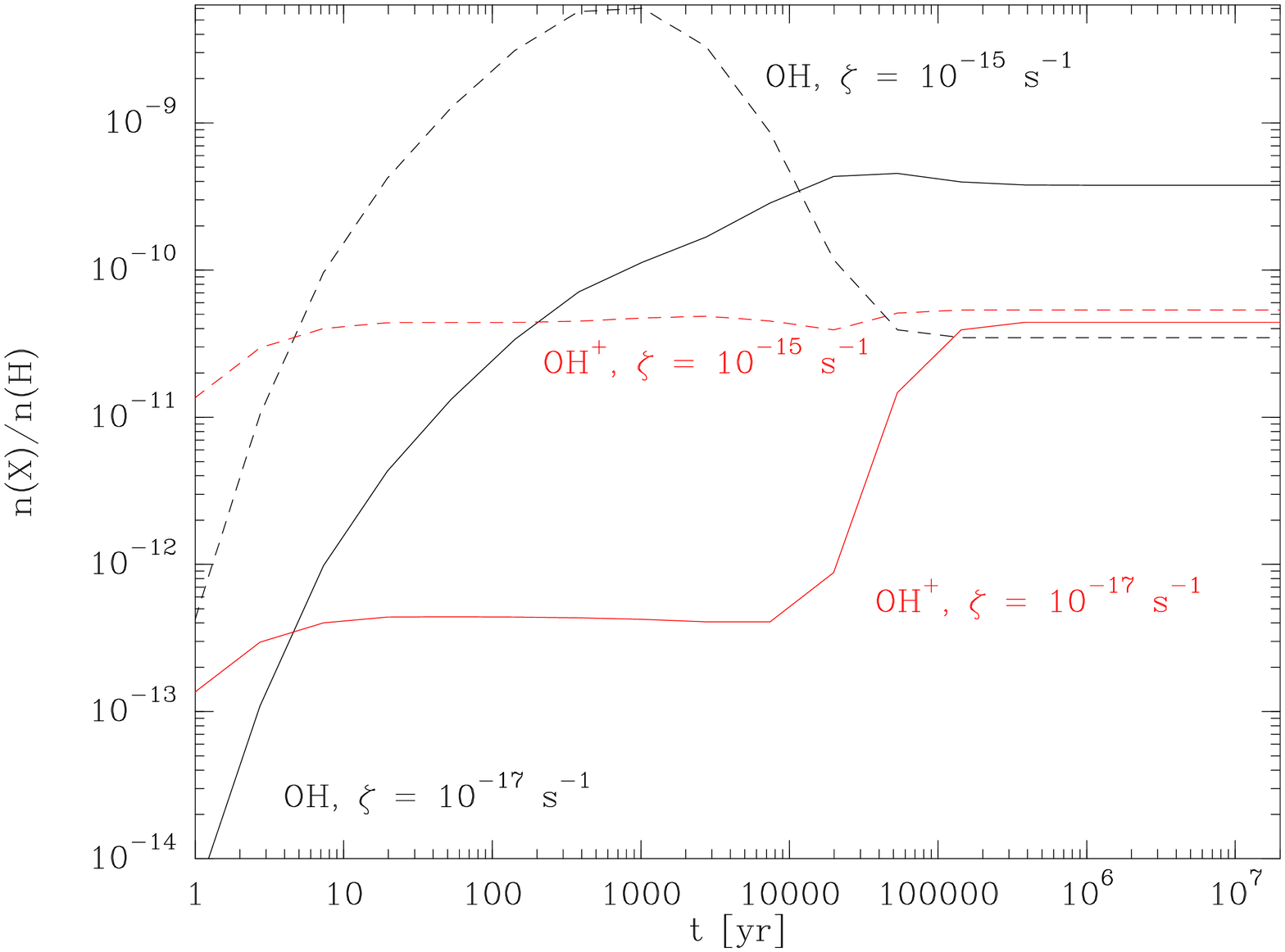}
   \includegraphics[width=\columnwidth]{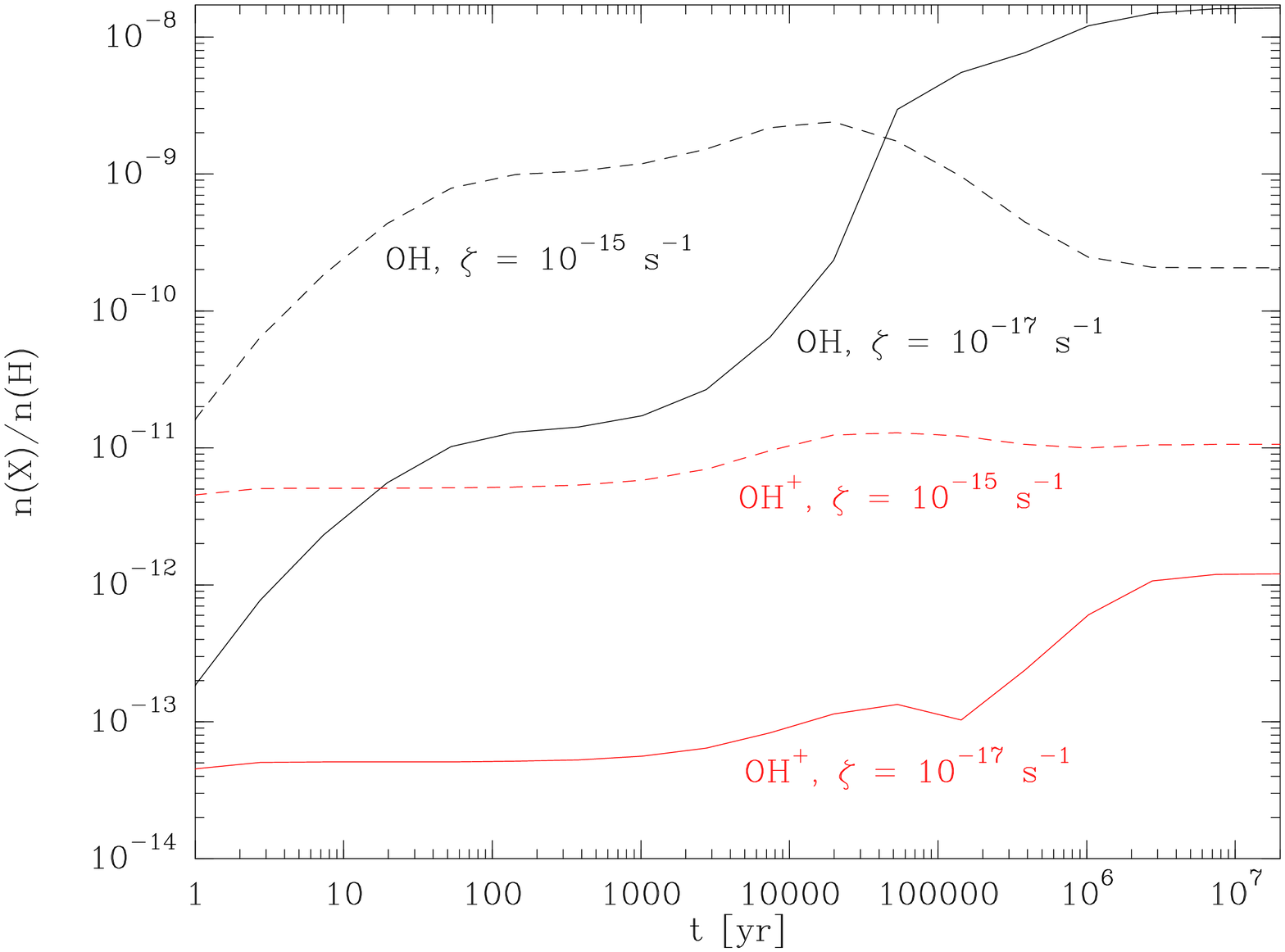}
   \caption{Chemical evolution for conditions typical of diffuse molecular (model 1, top) and translucent clouds (model 2, bottom). For details see
   text. Dashed curves refer to the case with enhanced cosmic ionization rate $\zeta = 10^{-15}\mathrm{s}^{-1}$.} 
   \label{fig:astrochem}
   \end{figure}
To estimate the production rate of OH and $\OHP$ we used a simple single-layer model. The purpose of this demonstration
is not a quantitative analysis, which would require dedicated PDR and TDR models \citep[][respectively]{1991ApJ...377..192H,2009A&A...495..847G}.
We used the Astrochem code (S. Maret, http://smaret.github.io/astrochem/) together with the reaction coefficients compiled by the KIDA database
(Wakelam et al. 2012, http://kida.obs.u-bordeaux1.fr). The models are defined in Table~\ref{table:2}. We used two values for the rate of ionization
by cosmic rays, $\zeta = 10^{-15}$ and $10^{-17}\,\mathrm{s}^{-1}$, and assumed the interstellar radiation field to be characterized by a typical
value of 1.7 Habing fields. The initial abundances with respect to H are 0.4 for $\HH$, 0.14 for He, and $3.2\times 10^{-4}$ for \OI.

The results are shown in Fig.~\ref{fig:astrochem}. During the evolution
toward equilibrium chemistry the \OI abundance remains constant at the initial value, except for the translucent cloud (model 2),
where it drops to $2\times 10^{-4}$ after $\sim$10$^5$~years because it is used for the enhanced production of OH and $\HHO$ thanks to the higher
$\HH$ fraction. However, for a translucent cloud exposed to the higher cosmic-ray ionization rate ($\zeta = 10^{-15}$), the \OI reservoir can be
replenished by the cosmic-ray-induced dissociation of OH and $\HHO$. In this case, the ratio $n(\OH)/n(\OHP)$ can reach values below unity.
For translucent clouds with their higher molecular hydrogen fraction, the conversion of $\OH^+$ to OH is much faster, as expected, with an $\OH/\OHP$
abundance ratio increased by up to several orders of magnitude. When chemical equilibrium is reached, the remaining column density of $\OHP$ is
too low to be detectable, leading to the observed peak in the contrast ratio shown in Fig.~\ref{fig:contrast}.

While these simple models qualitatively agree with our findings, a more realistic model would include not only a PDR and dissipation of turbulence,
but also a model for the evolution from the diffuse atomic gas to a molecular cloud. In this situation, which is beyond the scope of this paper,
the chemical equilibrium would be reached much later.
\section{Comparison of abundances of CH and HF}
\label{app:E}
To determine fractional abundances, we used HF and CH as surrogates for $\HH$
\citep{2009ApJ...706.1594N,2010A&A...521L..12S}. The interstellar chemistry of HF involves only four reactions and
can form in an exothermic reaction with $\HH$. 
\citet{2012A&A...540A..87G} predicted an HF abundance of $X$(HF)/$X(\HH) = 3.6\times 10^{-8}$, which is thought to hold for $A_{\rm V} > 0.2$
and $n_{\rm H} > 30$~cm$^{-3}$, but pointed out that this value needs to be confirmed by observations. Furthermore, fluorine forms in AGB stars,
at least in the solar neighborhood \citep{2014ApJ...789L..41J}, while it is destroyed in massive stars. It is therefore possible that fluorine
has a Galactic abundance gradient. \citet{2013ApJ...764..188I} have measured $X$(HF) using the $\upsilon=1-0, R(0)$ ro-vibrational
transition at $\lambda 2.5\mu$m and found values (e.g., $X(\HF) = 1.15\times 10^{-8}$ in the diffuse gas toward HD\,154368)
significantly below the prediction. \citet{2015ApJ...806...49S} confirmed these findings with models that yield $X(\HF) \sim 0.9\times 10^{-8}$ in
the low-density regime (at $A_{\rm V} \sim 0.9\times 10^{-8}$) to $X(\HF) \sim 3.3\times 10^{-8}$ at $A_{\rm V} \sim 4$~mag. The lower HF
abundance results from recent laboratory measurements \citep{2014NatCh...6..141T} that correct the reaction rate ${\rm F}(\HH,\H)\HF$ downward.
Using $X(\CH) = 3.5\times 10^{-8}$ \citep[UV spectroscopy]{2008ApJ...687.1075S}, \citet[further references therein]{2010A&A...521L..16G} showed
that their CCH abundance agrees with earlier findings. \citet{2012ApJ...756..136E} determined
$X(\HF) = 1.5\times 10^{-8}$ in the diffuse foreground gas toward massive star-forming regions, again using the canonical CH abundance
from \citet{2008ApJ...687.1075S}. We note that these measurements are fraught with uncertainties of typically 30 to 50\%. However, the overall
agreement, supported by the laboratory measurements, suggests that the $X(\CH) = 3.5\times 10^{-8}$ and $X(\HF)/X(\CH) = 0.4$ indeed lead to
a realistic HF abundance of $1.4\times 10^{-8}$. To substantiate these uncertainties, we used
CH in parallel wherever it was available. The large scatter in the ratio of $\HH$ column densities
derived from HF and CH (Table~\ref{tab:results}) reflects the underlying uncertainties. It seems likely, however, that the 
uncertainties in the assumed abundance of CH are larger. Its chemistry is more complex than that of HF \citep[e.g.,][]{1986ApJS...62..109V}
and its Galactocentric abundance gradient is known to be steeper than that of oxygen \citep[e.g.,][]{2005ApJ...618L..95E}.
The absorption of CH is not saturated, but the large hyperfine splitting (the components of the 536~GHz transition are separated by
19.3~\kms) and the fact that CH appears in emission at the velocities of the hot cores produce complex spectra. 
Our analysis can deal with this complexity: For the fits in Figs.~\ref{fig:spectra1stqu} and \ref{fig:spectra4thqu} only one
emission component with a given velocity, width, excitation temperature, and opacity had to be assumed.
Derived column densities are representative for $N_\HH$ \citep[cf][]{2002A&A...391..693L}. The 
abundances in Table~\ref{tab:results} derived from the column densities of both $\HH$ tracers agree reasonably well, although differences occur when HF tends to saturate.

\begin{figure}[ht!]
   \centering 
   \includegraphics[width=\columnwidth]{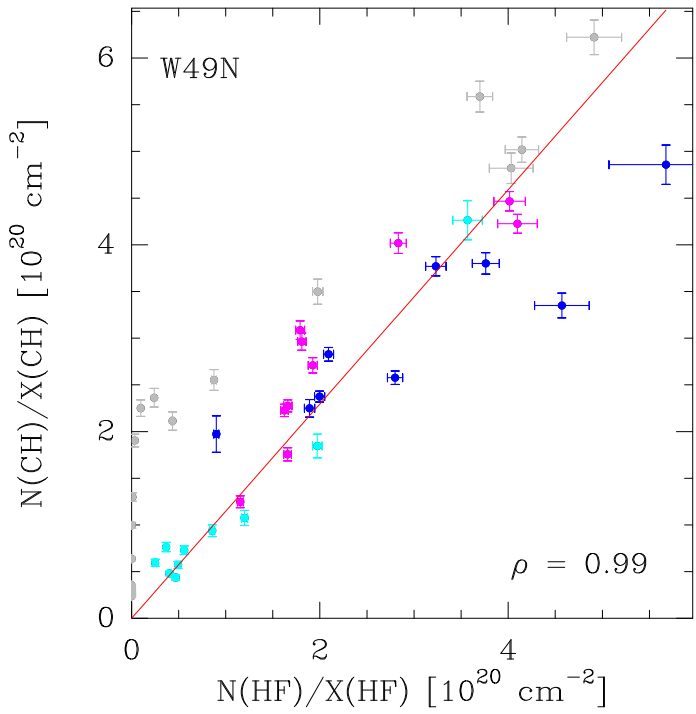}
   \caption{Comparison between $\HH$ column densities derived from CH and from HF toward W49N in velocity
            intervals of 1~\kms. The color code for the velocities is the same as in 
            Fig.~\ref{fig:oxygen_abundance} (velocities above 60~\kms are shown in gray). }
   \label{fig:compare_ch_hf}
   \end{figure}

For the example of W49N, Fig.~\ref{fig:compare_ch_hf} shows a comparison between the $\HH$ column densities derived from 
CH and those derived from HF. The correlation is statistically significant, with a false-alarm probability below 1\% and
with a correlation coefficient of 0.99. The slope of the regression line exceeds the expected slope of unity by 15\% and
corresponds to an X(HF)/X(CH) ratio of 0.35 instead of the canonical value 0.4, which is compatible with its standard deviation. 
Only velocities void of saturated absorption in HF and of emission in CH were considered ($>30$~\kms). Data points with 
$N/\sigma_{\rm N} < 5$ were discarded from this analysis. The result (slope and coefficient of the correlation) is fairly robust
against relaxing these restrictions within reasonable limits.

\section{Cross-calibration between PACS and GREAT}
\label{app:F}
\begin{figure}[ht!]
\centering
   \includegraphics[width=0.75\columnwidth]{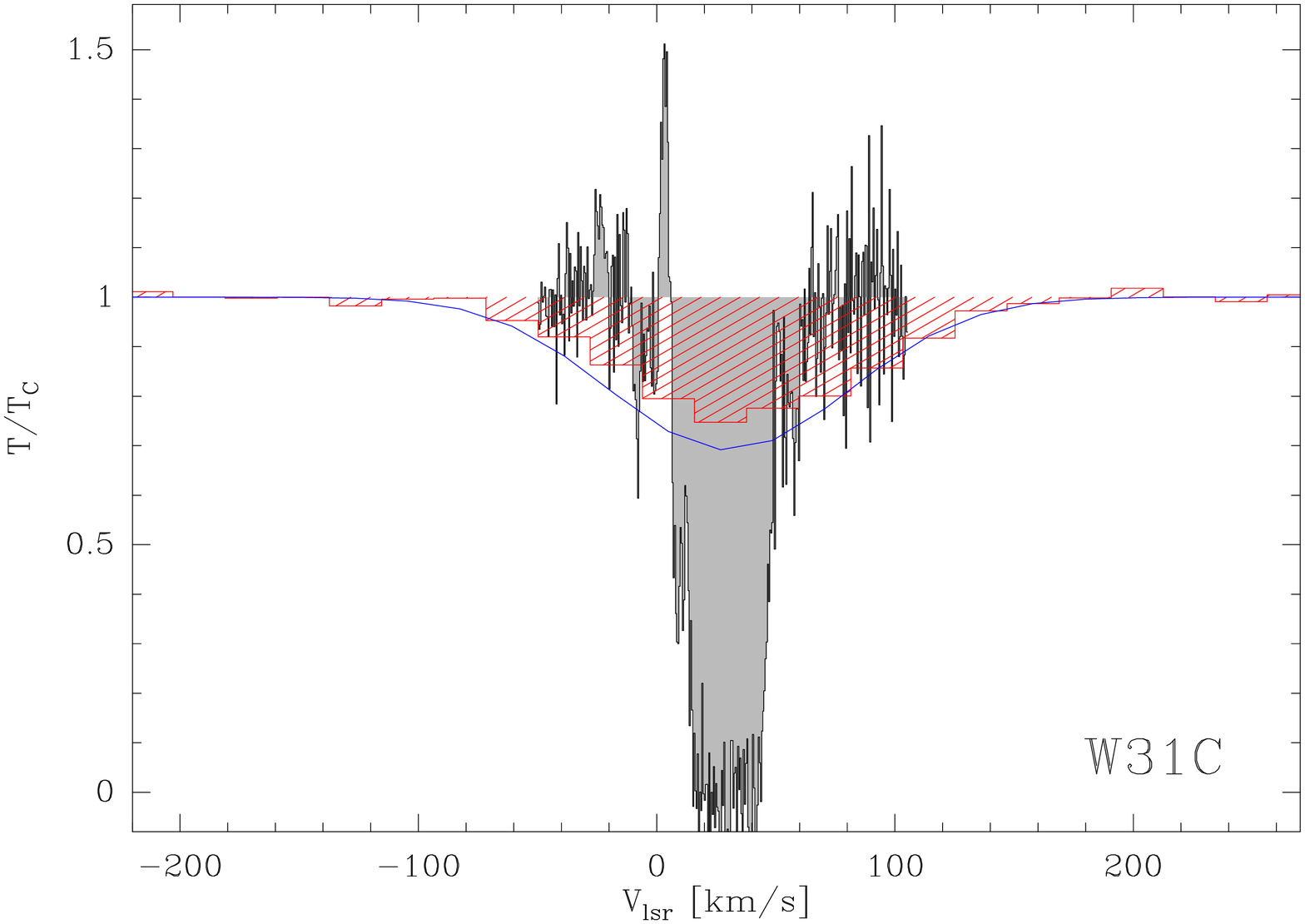}\vspace{2mm}
   \includegraphics[width=0.75\columnwidth]{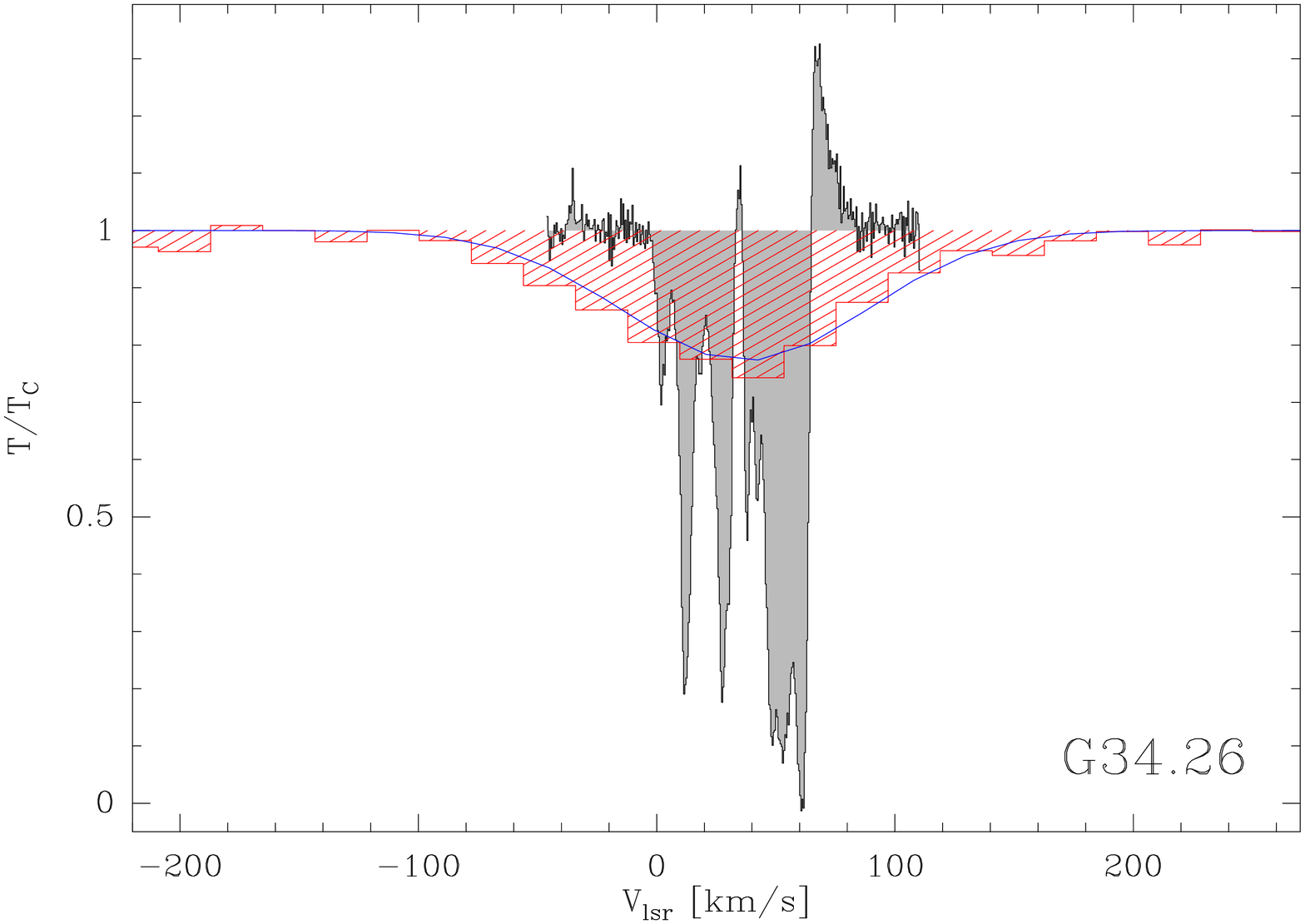}\vspace{2mm}
   \includegraphics[width=0.75\columnwidth]{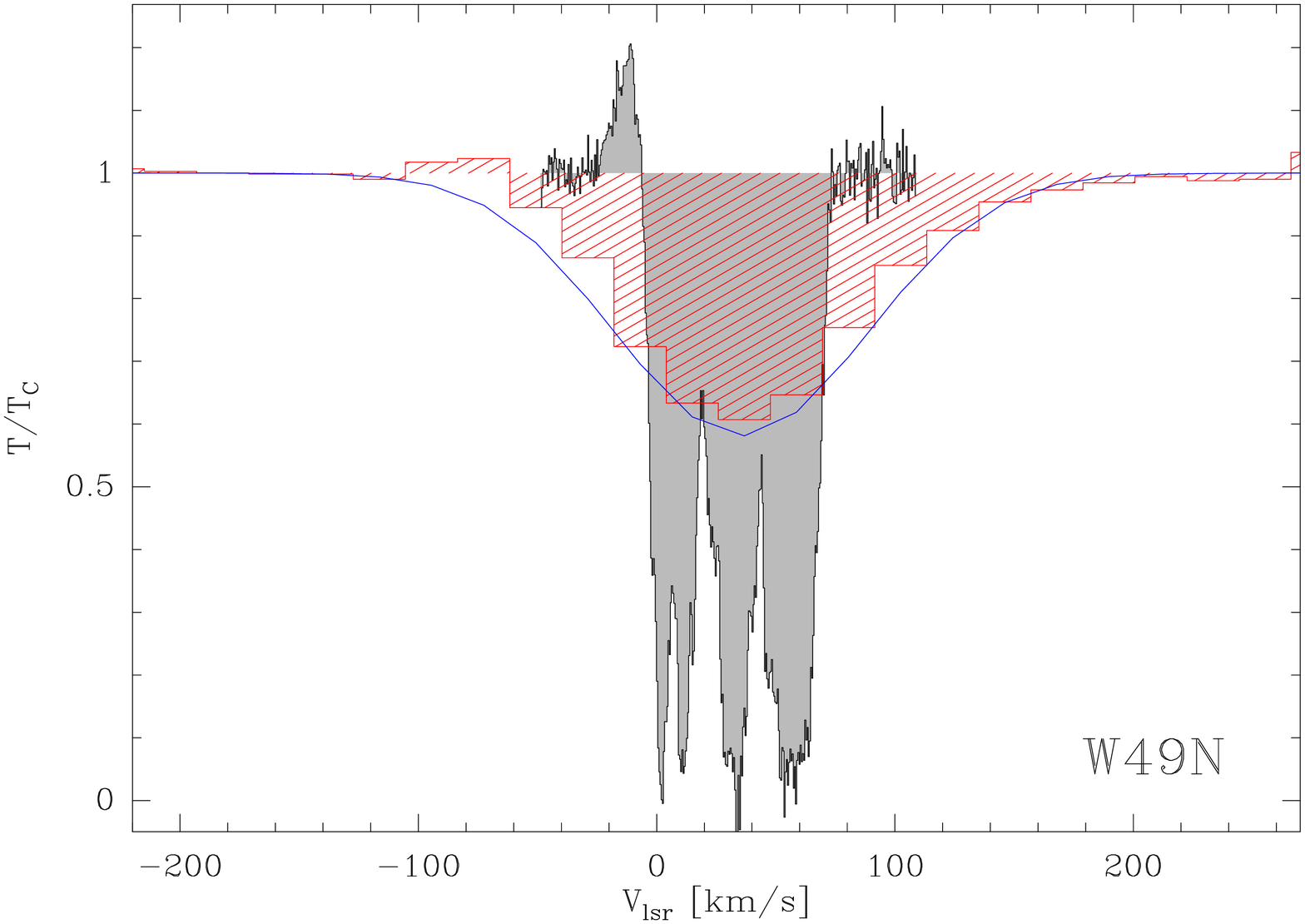}\vspace{2mm}
   \caption{Comparison between GREAT spectra of the \OI line (gray filled) with those observed with
   Herschel/PACS (red hatched). The blue spectrum results from the convolution of the GREAT spectrum
   with a Gaussian of 86.6~\kms width (FWHM). All spectra are normalized to unity at the continuum
   level. From top to bottom: W31C (G10.62), G34.26, and W49N.}
\label{fig:pacs}
\end{figure}

To confirm the continuum calibration in GREAT's  H-channel, here
we compare our flux densities with those measured by PACS. For the three targets of the \OI
observations, Herschel/PACS data exist (range spectroscopy mode B3A R1, Herschel key program ``PRISMAS'')
and were taken in pointed chop/nod mode. Figure~\ref{fig:pacs} shows a comparison between
the GREAT spectra and the \OI line as observed with PACS. We note that the large width of the spectral response of
PACS fills the saturated absorption with the continuum emission on either side of the line profile.
We convolved our high-resolution data with a Gaussian
profile with width 86.6~\kms (FWHM, from \citealp{2010A&A...518L...2P}, their Table~1), assumed to represent
the response of PACS to a spectrally unresolved feature. The skewness of the spectral response (PACS Observer's
Manual, Version 2.5.1) was not taken into account but may explain the difference between the PACS spectra and
our simulation. The continuum fluxes of PACS agree with ours (Table~\ref{tab:oxygen}) within 4 to 14\% (for 
G34.26 and W49N, respectively), and within 22\% for W31C.
\end{appendix}
\end{document}